\documentclass[review]{elsarticle}

\usepackage{rotating}
\usepackage{xcolor}
\usepackage{lineno,hyperref}
\usepackage{lscape}
\usepackage{placeins}
\modulolinenumbers[5]
\usepackage[margin=0.7 in]{geometry}
\usepackage{multicol}
\usepackage{verbatim}
\journal{Journal of \LaTeX\ Templates}

\begin{document}

\begin{frontmatter}

\title{Combining tumour response and progression free survival as surrogate endpoints for overall survival in advanced colorectal cancer } 

 \cortext[correspondingauthor]{Corresponding author}
 \emailauthor{eelia@hsph.harvard.edu}{Elia EG}

\author{Elia EG$^{1,2*}$, St{\"a}dler N$^{3}$, Ciani O$^{4,5}$, Taylor RS$^{4}$, Bujkiewicz S$^{1}$} 
\fntext[myfootnote]{Department of Health Sciences, University of Leicester, George Davies Centre, University Road, Leicester, LE1 7RH, UK}

\fntext[myfootnote]{Department of Biostatistics, Harvard University, 677 Huntington Ave, Boston, MA 02115, USA}

\fntext[myfootnote]{F. Hoffmann-La Roche Ltd, Basel, Switzerland }

\fntext[myfootnote]{Evidence Synthesis \& Modelling for Health Improvement, Institute of Health Research, University of Exeter Medical School, University of Exeter, Exeter EX2 4SG, UK }

\fntext[myfootnote]{ CERGAS Bocconi University, via Rontgen 1, 20136 Milan, Italy}


\begin{abstract}
Objectives:\\

\setlength{\parindent}{0cm}

To investigate whether joint modelling of two candidate surrogate endpoints, the tumour response (TR) and progression free survival (PFS), improves their predictive value of the final clinical endpoint, the overall survival (OS), in advanced colorectal cancer. \\

Study Design and Setting:\\

Data were obtained from a systematic review of randomised control trials investigating effectiveness of a broad range of pharmacological therapies in advanced colorectal cancer. The treatments included systemic chemotherapies, anti-epidermal growth factor receptor (anti-EGFR) therapies, anti-angiogenic agents, other multi-targeted antifolate (MTA) treatments and intra-hepatic arterial chemotherapy (IHA).
Multivariate meta-analysis was used to model the association patterns between treatment effects on the surrogate endpoints (PFS and TR) and the final outcome (OS). 
The surrogate relationships were explored by the use of regression parameters (intercept, slope and conditional variance) and adjusted R-squared.
Analyses were conducted on data from all trials reporting relevant outcomes and in subgroups defined by the type of therapy. \\

Results:\\

Analysis of 33 trials which reported treatment effects on all three outcomes (TR, PFS and OS) showed that there was association between treatment effects on OS and the treatment effect on both candidate surrogate endpoints, TR and PFS, when modelled individually and jointly. The surrogate association was particularly strong between treatment effect on PFS and OS, with the positive slope 0.28 (95\% CrI: 0.12, 0.45) and relatively small conditional variance 0.003 (95\% CrI: 0.00, 0.01). The adjusted R-square was also relatively high but obtained with a large uncertainty; 0.66 (0.2, 0.97). This relationship was not improved by inclusion of the treatment effect on the second surrogate endpoint, the TR.  There was no marked improvement in neither surrogacy patterns nor the precision of predicted treatment effect in the cross validation procedure when modelling the treatment effect on the two outcomes jointly. This was also the case for the subgroups of therapy, the sensitivity analysis including a larger set of data from trials reporting treatment effects on at least two outcomes and using a smaller set of data from the trials that did not allow for treatment crossover. There was a very modest improvement in the surrogacy pattern and the predictions obtained from the analysis of multiple surrogate endpoints in the subgroups of the anti-angiogenic agents.\\

Conclusion:\\

Overall, the joint use of two surrogate endpoints did not lead to improvement in the association between treatment effects on surrogate and final endpoints in advanced colorectal cancer. Only small improvement in the precision of the predicted treatment effects on the final outcome in studies investigating anti-angiogenic therapy was noted. However, the improvement was very modest and likely due to chance.\\
\end{abstract}


\end{frontmatter}



		
		
 
 	


\section{Introduction}

	Surrogate endpoints have been receiving increased attention by the research community in the last three decades as they offer a cost effective and quicker alternative to the use of final outcomes especially if they can be measured with a shorter follow-up period \cite{burzykowski2006evaluation}. For surrogate endpoints to be used effectively in clinical research, they need to be validated. There are three levels of surrogate endpoint validation: biological plausibility of association between outcomes, patient-level association between outcomes and study-level association \cite{taylor2009,ciani2017time}. For the purposes of this study we focus on the latter level of validation. Study level association is the hallmark of surrogacy, i.e. establishing whether the treatment effect on the surrogate endpoint is likely to predict a treatment effect on the clinical outcome. This is usually carried out through meta-analyses of randomised controlled trials (RCTs) and, in particular, using a bivariate meta-analysis \cite{daniels1997meta, buyse2000validation, bujkiewicz2015}.

To identify a surrogate endpoint for overall survival (OS) in advanced or metastatic colorectal cancer, a number of candidate endpoints have been investigated as potential surrogate endpoints, including progression free survival (PFS), tumour response (TR) or time to progression (TTP)  \cite{buyse2007progression, giessen2013progression, shi2015individual, ciani2015meta, chirila2012meta, tang2007surrogate}.
In previous work investigating trial-level surrogacy in advanced colorectal cancer, Buyse et al found that PFS was an acceptable surrogate endpoint to the overall survival \cite{buyse2007progression}.
A more recent study, investigating surrogacy patterns across a broader range of treatments in a meta-analysis of 101 RCTs, showed suboptimal validity of PFS as a surrogate endpoint for OS in advanced or metastatic colorectal cancer \cite{ciani2015meta}.
Other studies also investigated the surrogate relationship between the PFS and OS in advanced colorectal cancer that suggested that further validation is required \cite{giessen2013progression, shi2015individual}.

In these previous studies, using meta-analytic techniques either in the form of a meta-regression or a bivariate meta-analysis, only one surrogate endpoint at a time was investigated. Recently, researchers investigated use of multiple surrogate endpoints in a patient-level surrogate endpoint validation in multiple sclerosis \cite{sormani2009magnetic} and as joint predictors of clinical benefit measured on final outcome in a meta-analytic framework using Bayesian multi-variate meta-analysis \cite{bujkiewicz2015bayesian}.

In this paper we investigate whether the use of treatment effects on two candidate surrogate endpoints, PFS and TR, can result in increased precision of predicted treatment effect on the final outcome, namely OS, in advanced colorectal cancer when they are modelled jointly.
We investigated the predictive ability of the surrogate endpoints in this setting by conducting a bivariate meta-analysis to investigate one surrogate endpoint at a time and a trivariate meta-analysis to evaluate the surrogate endpoints jointly. 
We conducted the analysis in the Bayesian framework using multivariate meta-anlysis method described by Bujkiewicz et al \cite{bujkiewicz2015bayesian}. Then the added value of multiple surrogate endpoints modelled jointly was investigated, by comparing the meta-analytic models in terms of predicted intervals.

\section{Methods}

\subsection{Data sources}
\label{meth_data}
We used data from a systematic review by Ciani et al 2015 \cite{ciani2015meta}, which included treatment effect estimates from 101 randomised controlled trials (RCTs) in advanced or metastatic colorectal cancer that assess pharmacologic therapy against other therapy.
The studies included trials investigating a broad range of treatments including five different classes which were systemic chemotherapy, anti-epidermal growth factor receptor monoclonal antibodies (Anti-EGFR), anti-angiogenic agents, other multi-targeted antifolate (MTA) and intra-hepatic arterial chemotherapy (IHA).
For the purposes of the systematic review conducted by Ciani et al. \cite{ciani2015meta} the following definitions of outcomes were considered:
OS was defined as the time from randomization to time of death,
PFS was defined as the time from randomization to tumour progression (regardless of how the progression was defined), or death from any cause, and
TR was defined by objective tumor measurements by utilising methods that classify patients as responders, with a complete or partial
confirmed best response. Responses were determined using the criteria and recommendations according to the Response Evaluation
Criteria in Solid Tumors (RECiST) guidelines \cite{therasse2000new} or the World Health Organization recommendations \cite{world1979handbook}.\\

For purpose of this paper, we used data from the above systematic review on treatment effects measured on OS, PFS and TR.
Not all of the studies in the systematic review reported treatment effects on all three outcomes of interest.
In our base case scenario analysis, we used data from trials reporting all three outcomes. 
As a sensitivity analysis, we carried out an analysis on data from studies which reported treatment effects on at least two outcomes. Analyses were repeated in subsets of data defined by the type of treatment. A sensitivity analyses to examine the impact of an outlying observation and the choice of the prior distribution for the between-studies correlation were also carried out  (for details see section \ref{sec.stat.meth}).

Crossover in RCTs, for example from the control to experimental arm following progression, often results in loss of information about the treatment effect on the final outcome; what the effect would have been if crossover was not allowed. As patients move to the experimental treatment arm, the difference in treatment effect on OS between the treatment arms diminishes, leading potentially to zero effect with large uncertainty. This creates difficulty in estimating the association patterns between treatment effects on surrogate and final endpoint and the estimates of the latter are not reliable and their variability is reduced (potentially diminishing the correlation between the treatment effect on the two outcomes). Another sensitivity analysis was carried out on the subset of trials  which did not allow for crossover.

Individual patient data (IPD) were available from one of the RCT's included in the systematic review; study by Hurwitz et al. \cite{hurwitz2004bevacizumab} that investigated the use of bevacizumab in combination with irinotecan, fluorouracil, and leucovorin in patients with metastatic colorectal cancer \cite{hurwitz2004bevacizumab}. The IPD were used to obtain the within-study correlations between the treatment effects on the surrogate endpoints and on the final outcome. \\

\subsection{Statistical analysis}
\label{sec.stat.meth}
We used multivariate meta-analysis in a Bayesian framework to model jointly the treatment effects on one or two surrogate endpoints and on the overall survival which was the final clinical outcome. Treatment effects on PFS and OS were modeled using hazard ratios (HRs) and treatmet effects on TR were modelled using odds ratios (ORs). Log scale was used to allow the assumption of normality of the effects.
For studies, where there were no responders in one of the treatment arms, continuity correction of 0.5 was added to all values of the contingency table to enable finite odds ratio and variance estimators to be derived.

To model jointly treatment effect on two surrogate endpoints and the final outcome, we used trivariate random-effects model, where the estimates of treatment effects on the two surrogate endpoints, log OR of TR, denoted $Y_{1i}$ and log HR on PFS denoted $Y_{2i}$ and the treatment effect on the final outcome; log HR on OS, denoted $Y_{3i}$ are assumed to be correlated and normally distributed
\begin{equation}
	\left(
	\begin{array}{c}
	Y_{1i}\\
	Y_{2i}\\
	Y_{3i}\\
	\end{array}
	\right)\sim
	N\left(
	\left(
	\begin{array}{c}
	\mu_{1i}\\
	\mu_{2i}\\
	\mu_{3i}\\
	\end{array}
	\right), \mathbf{\Sigma_i}
	\right),\;
	\mathbf{\Sigma_i} =
	\left(
	\begin{array}{ccccc}
	\sigma_{1i}^2 & \sigma_{1i}\sigma_{2i} \rho_{wi}^{12}  & \sigma_{1i}\sigma_{3i} \rho_{wi}^{13}  \\
	\sigma_{2i}\sigma_{1i}\rho_{wi}^{12} & \sigma_{2i}^2 & \sigma_{2i}\sigma_{3i} \rho_{wi}^{23}  \\
	\sigma_{3i}\sigma_{1i}\rho_{wi}^{13} & \sigma_{3i}\sigma_{2i}\rho_{wi}^{23} & \sigma_{3i}^2 \\
	\end{array}
	\right) \nonumber ,
\label{wsm}
\end{equation}
The  trivariate distributions describe the within-study variability, with the effects $Y_{ki}$ estimating the true treatment effects $\mu_{ki}$ in the population and $\sigma_{ki}^2$ are the corresponding variances of the estimates of the treatment effects on outcome $k=1, \dots, 3$  in each study $i$ and $\rho_{wi}^{kl}$ the within-study correlations between these estimates. The elements of the within-study variance-covariance matrix are assumed known.
To describe the between-study variability we modelled the correlated true treatment effect $\mu_{ki}$ in  the product normal formulation of conditional univariate normal distributions:
\begin{equation}
		\begin{array}{c}
		\mu_{1i} \sim N(\eta_{1},\psi_1^2)\\
		\mu_{2i} \mid \mu_{1i} \sim N(\eta_{2i},\psi_2^2)\\
		 \eta_{2i}=\lambda_{20} + \lambda_{21} \; \mu_{1i}\\
		\mu_{3i} \mid \mu_{2i} \sim N(\eta_{3i},\psi_3^2)\\
	 \eta_{3i}=\lambda_{30} + \lambda_{32} \; \mu_{2i}
	 	\end{array}
\label{bsm}
\end{equation}
where the variances $\psi_k^2$, $k=1,2,3$ relate to the between-study heterogeneity parameters $\tau_{k}^{2}$;   $\psi_{1}^{2}=\tau_{1}^{2}$, $\psi_{2}^{2}=\tau_{2}^{2}-\lambda_{21}^{2}\tau_{1}^{2}$ and $\psi_{3}^{2}=\tau_{3}^{2}-\lambda_{32}^{2}\tau_{2}^{2}$, with the regression coefficients related to both the heterogeneity parameters and the between-studies correlations $\rho_b^{kl}$: $\lambda_{21}=\rho_{b}^{12}\frac{\tau_{2}}{\tau_{1}}$, $\lambda_{32}=\rho_{b}^{23}\frac{\tau_{3}}{\tau_{2}}$. In the Bayesian framework the parameters are given prior distributions.
The between-studies correlations are given informative prior distributions as recommended by Burke et al \cite{burke2016}. 
By assuming that an increase in the treatment effect on PFS (reduced progression rate) will lead to an increased effect on OS (reduced mortality rate) and hence a positive correlation, we place a uniform prior distribution allowing only positive values between zero and one for the correlation is used; $\rho_{b}^{23}\sim U(0,1)$.
In a similar manner, assuming that increased response rate would lead to reduced progression or mortality rates, and hence a negative correlation, we place prior distribution that allows only negative values between minus one and zero on the correlations between treatment effects on TR and PFS and between the effects on TR and OS;   $\rho_{b}^{12},\;\; \rho_{b}^{13}\sim U(-1,0)$. Sensitivity analysis was carried out using a non-informative prior distribution U(-1,1) for between-studies correlation.
The heterogeneity parameters are given half normal distributions, $\tau_{1,2,3}\sim N(0, 1000)I(0,)$, to allow only for positive values  \cite{bujkiewicz2015}. The remaining parameters are given non-informative normal prior distributions; $\eta_{1}\sim N(0, 1000)$, $\lambda_{20},\;\;\lambda_{30}\sim N(0, 1000)$.
	
	In the above model describing the between-studies variability, we used structured between-studies variance-covariance matrix by assuming conditional independence between the true treatment effects on the first surrogate endpoint $\mu_{1i}$ and on the  final outcome $\mu_{3i}$. 
As a sensitivity analysis we also used an alternative model using unstructured variance-covariance matrix and another model with a different structure on the variance-covariance matrix where the true treatment effects on the two surrogate endpoints, $\mu_{1i}$ and $\mu_{2i}$, were assumed conditionally independent. 
	
As stated above,  the within-study correlations are assumed to be known. However, to obtain the correlations, IPD are necessary as the correlations between treatment effects on log HR and log OR scales are not reported in the original RCT reports.
For the purposes of this study, IPD from only one study were available \cite{hurwitz2004bevacizumab}. Within-study correlations between the treatment effects on the three outcomes (TR, PFS and OS) were obtained using bootstrapping methods with 5000 bootstrap samples \cite{efron1992bootstrap}. We assume that the within-study correlations were equal across all studies.

Two models were considered, the bivariate and the trivariate random effects models. The bivariate model was used to describe and evaluate the association between treatment effects on a single surrogate endpoint (PFS or TR) and on the final outcome (OS), whilst the trivariate model was used to describe
 the association between treatment effects on multiple surrogates (PFS and TR) jointly and treatment effect on the final outcome (OS).
 The trivariate model described by the formulae (\ref{wsm})-(\ref{bsm}) reduces to bivariate meta-analytic model when there are only two outcomes.
More details on product normal formulation of the bivariate  model for surrogate endpoints can be found in Bujkiewicz et al. \cite{bujkiewicz2015} and for the trivariate (and multivariate)  model for multiple surrogate endpoints in Bujkiewicz et al.  \cite{bujkiewicz2015bayesian} and for borrowing of strength across outcomes with missing data in Bujkiewicz et al \cite{bujkiewicz2013multivariate}. \\

\subsubsection{Surrogacy relationship evaluation}
	We followed the surrogacy criteria introduced by Daniels and Hughes \cite{daniels1997meta}, and adopted by Bujkiewicz et al.  \cite{bujkiewicz2015bayesian}, by which the slope in (\ref{bsm}), $\lambda_{32}$, indicates the association between the treatment effect on the second surrogate endpoint (PFS) and the treatment effect on the final outcome (OS).
For the treatment effects to be associated,  we require the slope $\lambda_{32}\neq 0$.
For the association to be perfect the conditional variance $\psi^{2}_{3}=0$.
Moreover, we would expect the intercept $\lambda_{30}=0$ to ensure that no treatment effect on surrogate endpoint will imply no treatment effect on the final outcome.
In a similar manner we can describe surrogacy criteria between the first and second surrogate; $\lambda_{21}\neq 0$, $\psi^{2}_{2}=0$ and $\lambda_{20}=0$, where the second surrogate becomes the final outcome in a bivariate model with a single surrogate endpoint. \\
We also report the adjusted $R^{2}$ \cite{renfro2012bayesian, burzykowski2001validation} which for perfect surrogacy relationship should be one. In our model $R_{adj, 23}^{2}=\lambda_{32}^{2}\frac{\tau_{2}^{2}}{\tau_{3}^{2}}$ for the relationship between treatment effects on the second surrogate endpoint and on the final outcome.\\

\subsubsection{Cross validation and model comparison}

In order to investigate whether the joint use of treatment effects on multiple surrogate endpoints gives more precise predictions of the treatment effect on the final outcome, a cross-validation procedure was carried out. In one study at a time the treatment effect on the final outcome was assumed unknown and predicted from the treatment effect on surrogate endpoint (or multiple surrogate endpoints jointly) using the bivariate (or trivariate) meta-analytic model.
In this Bayesian approach to multivariate meta-analysis, this was achieved by assuming that the unreported outcomes were missing at random,
which were then predicted by the Markov chain Monte Carlo (MCMC) simulation of the model \cite{bujkiewicz2013multivariate}.
The predicted interval was obtained by assuming the variance $\sigma_{3i}$ (or $\sigma_{2i}$ in the bivariate case) of the treatment effect on the final outcome known and inflating it by the variance of the random effect giving the variance of the predicted effect
$\sigma_{3n}^2+var(\hat{\mu}_{3n}\vert Y_{1n},Y_{2n},\sigma_{1n},\sigma_{2n},Y_{1(-n)},Y_{2(-n)},Y_{3(-n)})$,
where $Y_{1(2,3)(-n)}$ denote the data from the remaining studies without the validation study $n$, similarly as in Daniels and Hughes \cite{daniels1997meta}.
The added value of multiple surrogate endpoints modelled jointly was then investigated by comparing the parameters describing criteria for surrogacy (slope, intercept and conditional variance) and the predicted effects obtained from the cross-validation procedure. The predicted effects on the final outcome obtained by the bivariate and the trivariate meta-analytic models were compared in terms of the width of the  predicted intervals. We investigated surrogacy across all RCTs as well as in subgroups of class of therapy.

\subsection{Software and computing}

All models were implemented in WinBUGS \cite{Lunn2000} where the estimates were obtained using MCMC
simulation using 250000 iterations (including 150000 burn-in).
Convergence was checked by visually assessing the history, chains and autocorrelation using graphical tools in WinBUGS. 
All posterior estimates are presented as means with the 95\% credible intervals (CrI).
R was used for data manipulation and to execute WinBUGS code multiple times (for validation of surrogates for each study) using the
R2WinBUGS package \cite{R2w}. OpenBUGS and R2OpenBUGS version of the software was used for the
cross-validation procedures which were conducted using Linux (Red Hat, Inc., Raleigh, North Carolina)-based high
performance computer.
WinBUGS programs corresponding to the bivariate and trivariate models are included in the online appendix.

\section{Results}
\subsection{Included data}
Out of the 101 RCT's 99 reported estimates of treatment effects on at least one of the three outcomes, 51 studies reported the treatment effects on at least two of the outcomes and 33 studies reported treatment effects on all three outcomes.
In the base case analysis of data from studies reporting all three outcomes (33 studies), subgroups of therapy included 15 studies which investigated the use of systemic chemotherapy, eight studies which evaluated the use of anti-EGFR therapies, nine studies investigating the use of anti-angiogenic agents and one study evaluating the use of IHA.
More details about the study characteristics are included in Ciani et al \cite{ciani2015meta}.

In the sensitivity analysis of studies reporting at least two outcomes, we used 51 of which: 48 studies invastigated OS (23 Systemic Chemotherapy,  11 anti-EGFR, 12 anti-angiogenic agents, 1 MTA, 1 IHA), 39 studies investigated PFS (17 Systemic Chemotherapy, 9 anti-EGFR,  12 anti-angiogenic agents, and 1 IHA), and 48 studies investigated TR (25 Systemic Chemotherapy, 10 anti-EGFR, 11 anti-angiogenic agents, 1 MTA, 1 IHA). 
The results obtained from the trivariate meta-analysis, where at least two outcomes were reported (with some missing data), were compared to the bivariate analysis of the complete data for each surrogacy pair:
 45 studies investigated both TR and OS  (23 Systemic Chemotherapy, 10 anti-EGFR, 10 anti-angiogenic agents, 1 MTA and 1 IHA);   
36 studies investigated TR abd PFS  (17 Systemic Chemotherapy, 8 anti-EGFR, 10 Anti-angiogenic agents, 1 IHA);
36 studies investigated  PFS  and OS (15 Systemic Chemotherapy, 9 anti-EGFR, 11 Anti-angiogenic agents and  1 IHA).

In the sensitivity analysis of trials that did not allow for patient crossover we combined seven studies, out of the total 33 trials reporting all three outcomes. The studies included three trials of systemic chemotherapy, one anti-EGFR therapy, two anti-angiogenic agents and one IHA.\\
List of references for studies included in the analysis can be fount in the supplementary materials.\\
Exploratory analysis of the data (presented graphically in Figures 1 to 4 and Table 1 of the supplementary materials) showed a lot of heterogeneity of the treatment effects for TR and PFS, with the confidence intervals of the treatment effects on TR particularly wide, especially for two classes of therapy: the chemotherapy and anti-EGFR therapies.\\

\subsection{Within-study correlations}
The within-study correlations required to populate the within-study variance covariance matrix in the multivariate meta-analytic model, were obtained from IPD. The within-study correlations between each pair of treatment effects on the three endpoints are as follows: the correlation between treatment effects on PFS and OS (log HR(PFS) and log HR(OS)) is $\rho_{wi}^{32}$= 0.513, correlation between treatment effects of TR and OS (log OR(TR) and log HR(OS)) is $\rho_{wi}^{31}$= -0.333 and the correlation between treatment effects on TR and PFS  ((log OR(TR) and log HR(PFS)) is $\rho_{wi}^{21}$= -0.433.

\subsection{Surrogacy criteria: base case scenario}
Three bivariate meta-analyses were carried out investigating the association between treatment effect on each pair of the three outcomes: TR, PFS and OS. Treatment effect on TR was evaluated as a surrogate to the treatment effect on PFS and to treatment effect on OS and the treatment effect on PFS was investigated  as surrogate to treatment effect on OS. Trivariate model was then used to investigate whether treatment effects on both TR and PFS modelled jointly improved their predictive value of the treatment effect on OS. Table \ref{bivcom} shows results of these analyses conducted in the base case  scenario where the data used were from studies reporting treatment effects on all of the three outcomes.

\begin{table}[h]
	\centering
	\scalebox{0.8}{
	\begin{tabular}{lccccc}
	   &TR OS&TR PFS & TR PFS & PFS OS  & PFS OS                  \\
	&2D & 2D&3D &2D & 3D\\	
		\hline

\multicolumn{6}{l}{\emph{All treatments (N=33)}} \\
intercept&-0.03(-0.07,0.02)&-0.05(-0.14,0.02)&-0.05(-0.13,0.02)&-0.02(-0.06,0.03)&-0.02(-0.06,0.03)\\
slope&-0.05(-0.13,0)&-0.32(-0.45,-0.2)&-0.31(-0.43,-0.19)&0.22(0.03,0.41)&0.19(0.02,0.4)\\
variance&0(0,0.01)&0.02(0.01,0.05)&0.02(0.01,0.04)&0(0,0.01)&0(0,0.01)\\
$R^{2}_{adjusted}$&0.33(0,0.91)&0.61(0.27,0.87)&0.64(0.3,0.89)&0.58(0.06,0.97)&0.5(0.02,0.95)\\

\multicolumn{6}{l}{\emph{Systemic chemotherapy (N=15)}} \\
intercept&-0.02(-0.08,0.04)&-0.04(-0.16,0.06)&-0.04(-0.14,0.05)&-0.02(-0.08,0.04)&-0.02(-0.08,0.04)\\
slope&-0.03(-0.11,0)&-0.26(-0.42,-0.08)&-0.25(-0.4,-0.09)&0.17(0,0.45)&0.14(0,0.4)\\
variance&0(0,0.01)&0.02(0,0.08)&0.02(0,0.06)&0(0,0.01)&0(0,0.01)\\
$R^{2}_{adjusted}$&0.39(0,0.96)&0.58(0.07,0.96)&0.66(0.11,0.98)&0.52(0.01,0.98)&0.47(0,0.97)\\

\multicolumn{6}{l}{\emph{Anti-EGFR  therapies  (N=8)}} \\
intercept&-0.06(-0.16,0.09)&-0.21(-0.37,0.01)&-0.18(-0.37,0.05)&-0.06(-0.16,0.14)&-0.04(-0.16,0.19)\\
slope&-0.04(-0.18,0)&-0.14(-0.36,-0.01)&-0.16(-0.39,-0.02)&0.14(0,0.63)&0.17(0,0.78)\\
variance&0.01(0,0.05)&0.02(0,0.08)&0.02(0,0.1)&0.01(0,0.04)&0.01(0,0.05)\\
$R^{2}_{adjusted}$&0.19(0,0.82)&0.45(0,0.96)&0.5(0.01,0.97)&0.22(0,0.85)&0.19(0,0.83)\\

\multicolumn{6}{l}{\emph{anti-angiogenic agents  (N=9)}} \\
intercept&0.04(-0.09,0.2)&0.03(-0.18,0.25)&0.03(-0.18,0.25)&0.02(-0.1,0.15)&0.02(-0.09,0.14)\\
slope&-0.35(-0.89,-0.03)&-0.87(-1.64,-0.3)&-0.85(-1.64,-0.27)&0.38(0.05,0.79)&0.37(0.04,0.77)\\
variance&0.02(0,0.07)&0.04(0,0.15)&0.03(0,0.14)&0.02(0,0.06)&0.01(0,0.06)\\
$R^{2}_{adjusted}$&0.52(0.01,0.97)&0.74(0.13,0.99)&0.72(0.1,0.99)&0.59(0.03,0.97)&0.56(0.02,0.97)\\

\hline	
\end{tabular}}
\caption{Surrogacy criteria obtained from bivariate (2D) and trivariate (3D) models for the association between treatment effect on surrogate (TR or PFS) and final outcome (OS and PFS in bivariate case). The results are posterior means and 95\% credible intervals.
TR -- tumor response, PFS -- progression free survival, OS -- overall survival}
	\label{bivcom}
\end{table}

Results of the three bivariate models applied to all of the data, presented in the top part of Table \ref{bivcom}, showed that there was an association between the treatment effects on each pair of outcomes. The intervals of the intercepts obtained from the bivariate models all contained zero indicating that no effect on the surrogate endpoint could imply no effect on the final outcome. The intervals for the slopes did not contain zero indicating positive association were slope was positive and negative association where slope was negative. However the surrogate relationships were not strong.  
When investigating TR as a surrogate endpoint for OS, the association between the treatment effects on the two outcomes was weak in terms of the small slope $\lambda_{21}$= -0.05 (95\%CrI:  -0.13,0.00) and the 95\% CrI contained zero when rounded to the second decimal place, and the mean and the lower bound of $R^{2}_{adj}=$ 0.33 (95\% CrI:  0.00, 0.91) were also small. 
The slope and the adjusted R-squared were higher for the relationship between the treatment effects on PFS and TR; slope was -0.32 (95\% CrI:  -0.45, -0.20)  and $R^{2}_{adj}=$ 0.61 (95\% CrI: 0.27, 0.87). 
However the lower bound of the CrI for the R-squared was still low and the conditional variance was relatively high,  0.02 (95\% CrI: 0.01, 0.05), indicating a weak surrogate relationship. 
The surrogate relationship between the treatment effects on PFS and OS appeared stronger in terms of the conditional variance 0.00 (95\% CrI: 0.00, 0.01) and the $R^{2}_{adj}=$ 0.58 (95\% CrI: 0.06,0.97) with lower ends of CrIs being close to zero.

Results from the trivariate meta-analysis, which described the associations between the treatment effects on the two pairs of outcomes (effects on PFS and TR and effects on OS and PFS) in a single model, were similar to those obtained from separate bivariate models. Precision around the intercept, slope and the conditional variance was minimally reduced for the association between the treatment effects on PFS and TR in the trivariate analysis whereas precision for these estimates for the association between the treatment effects on OS and PFS remained the same in the trivariate analysis as in the the bivariate analysis using a single surrogate endpoint.

Table \ref{bivcom} also shows the results for each subclass of therapy. For subgroup of trials investigating systemic chemotherapy, the results were similar to those obtained from the whole cohort of studies but typically obtained with increased uncertainty (wider CrIs) and weaker association in terms of lower mean slope. The adjusted R-squared was minimally higher in this subgroup for the association between the treatment effects on OS and TR, whilst a lower mean slope was obtained for the association between TR and PFS and between PFS and OS, compared to the analysis of all treatments. 
The slopes for the association between the treatment effects on TR and PFS and between the effects on OS and PFS were obtained with minimally higher precision from the trivariate models compared to the bivariate.
For the anti-EGFR therapies, also similar results were obtained to those from the analysis conducted on data from all of the studies but also, similarly as for systemic chemotherapy, with weaker association pattern.
For anti-angiogenic agents the mean slopes and the mean R-squared values were considerably higher for all investigated surrogacy relationships compared to other subclasses and the analysis of all treatments, however they were obtained with high uncertainty, also likely due to the small number of studies in the subgroup. 

\subsection{Cross-validation}
To investigate the  predictive value of the surrogate endpoints when modelled jointly, a cross-validation procedures were carried out.
The predicted treatment effects on OS predicted from the treatment effect on PFS alone were compared with those predicted from treatment effects on both TR and PFS jointly by exploring the associated uncertainty described by the predicted intervals as well as whether the predicted intervals contained the observed point estimate of the treatment effect on OS in each study.
When looking at predicted treatment effects obtained using the complete data set of the 33 studies, some of the predicted intervals were inflated when making predictions from the treatment effect on both surrogate endpoints jointly, compared to the predictions made from the treatment effect on PFS only. The intervals were on average 0.21\% wider with the percentage change of the width of the interval ranging between 1.96\% reduction to 4.4\% increase.  However, from the point of view of the cross-validation procedure, the 95\% predicted intervals included the observed point estimate in most of the studies apart from one which was an extreme lowest value of the treatment effect on OS. This was the case for both bivariate and trivariate models. Full set of predicted intervals is included in Table 2 of the supplementary materials.

The cross-validation procedure was repeated for each class of therapy separately. In terms of predicted intervals, on average there was a very modest improvement in precision of predictions for the anti-angiogenic agents obtained from the trivariate compared to the bivariate model (on average 2.35\% narrower 95\% predicted interval).  For the systemic chemotherapy class there was a minimal improvement in precision for most of the studies, with an average percentage reduction of 1.41\% predicted interval. However, the predicted intervals were inflated for the anti-EGFR therapies with, on average, 6.9\% increase in the width of the interval. All predicted intervals are included in Table 3 of the supplementary materials.

Overall there was not much benefit of combining treatment effects on two surrogate endpoints to predict the treatment effect on the final outcome. 
This lack of improvement, or even increased uncertainty of the predicted effect when using multiple surrogate endpoints,
 may be due to increased overall between-studies heterogeneity when extending the data to include the treatment effect on TR. 
 The between-studies heterogeneity for the treatment effect on TR was considerably higher compared to the heterogeneity of the treatment effects on PFS and OS in the data set including all treatments. 
 This was also the case for the subgroups of studies including the systemic chemotherapy  and the anti-EGFR therapy trials. 
 However, for the anti-angiogenic agents, the between-studies heterogeneity of the treatment effect on TR was comparable with that for the treatment effects on PFS. This may explain some increase in precision of the slope and the predicted effects on OS when using multiple surrogate endpoints in this class of therapy, as including additional outcome did not increase overall uncertainty.
  However, due to small number of studies the added value was minimal. The treatment effects on all three outcomes are comparable between those from bivariate and trivariate models. All mean treatment effects and the heterogeneity parameters are listed in Table 1 of the supplementary materials.
 
\subsection{Sensitivity analysis}
Sensitivity analysis, extending the data set to the 51 studies  reporting at least two outcomes gave similar results; surrogacy criteria for the association between the treatment effects on PFS and OS were satisfied both when looking at all therapies, and for Systemic chemotherapy and Anti-angiogenic agents therapies (Tables 4 and 5 in the online Appendix).

Results of the sensitivity analysis of trials with no crossover, presented in Table 2, were similar to those obtained from base case scenario with respect to the surrogacy criteria. The treatment effects on the surrogate and final outcomes appeared to be associated for all investigated surrogacy relationships. 
However for all sets of surrogacy relationships and both the bivariate and the trivariate analyses all estimates were obtained with large uncertainty compared to the whole set of studies, which was most likely due to the small number of studies (only 7) in the meta-analysis.
The association between treatment effects on OS and PFS appeared to be weaker in terms of the conditional variance which increased compared to the results obtained from all 33 studies reporting all three outcomes. This was the case in both sets of results, from the bivariate and the trivariate analyses.
This also led to reduced values of the adjusted R-squared.
The use of multiple surrogate endpoints did not improve the strength of the association patterns when compared to the use of a single surrogate endpoint in this subset of studies.
Similarly as for the full data set, this could also be due to the large heterogeneity  of the treatment effects on TR (see table 6 in the online appendix, including the average effects and the heterogeneity parameters.

The cross validation showed on average a very modest increase in precision, by on average 2.3\% and up to 6.2\%, when predicting the treatment effect on OS from the treatment effects on both PFS and TR compared to the predictions made from the effect on PFS only. Full set of results from the cross validation are included in Table 7 of the online appendix.

An additional sensitivity analysis was carried out to investigate an impact of an outlying observation 
(study with the largest effect size estimate for TR). The results (listed in Table 8 of the online appendix) showed an increased mean slope and the mean R-squared for the full set of studies and the subgroups of the EGFR inhibitors, however the uncertainty around these parameters in both sets of results also increased.\\

A final sensitivity analysis was carried out investigating the impact of the choice of the prior distribution for the between-studies correlation, replacing the informative prior distributions U(-1,0) for the negative association and U(0,1) for the positive association with a non-informative prior distribution U(-1,1). The results of this analysis were somewhat different compared to those obtained from the main analysis and are listed in Table 9 of the online appendix. The association between PFS and OS, and also for TR and PFS was satisfied for both 2D and 3D models applied to all treatments in the base case scenario. In terms of subclass therapies the association was preserved between TR and PFS for: Systemic chemotherapy and Anti-angiogenic agents.\\

\begin{table}[h!]
\centering
\begin{tabular}{ccccccc}	
		&TR OS&TR PFS & TR PFS &PFS OS &PFS OS                                       \\
		&2D &2D & 3D &2D & 3D\\	
		\hline
		
 intercept&-0.04(-0.15,0.09)&-0.05(-0.2,0.11)&-0.05(-0.2,0.1)&-0.03(-0.14,0.11)&-0.03(-0.15,0.1)\\
 slope&-0.1(-0.32,0)&-0.31(-0.64,-0.05)&-0.31(-0.64,-0.05)&0.28(0.01,0.88)&0.27(0,0.88)\\
 variance&0.01(0,0.04)&0.02(0,0.08)&0.02(0,0.08)&0.01(0,0.04)&0.01(0,0.04)\\
$R^{2}_{adjusted}$&0.4(0,0.96)&0.63(0.03,0.99)&0.64(0.03,0.99)&0.42(0,0.97)&0.4(0,0.96)\\

		\hline	
\end{tabular}
\caption{Surrogacy criteria obtained from the bivariate models and trivariate model applied to the trials that did not allow for the treatment crossover.
TR -- tumor response, PFS -- progression free survival, OS -- overall survival, 2D -- bivariate model using a single surrogate endpoint, 3D -- trivariate model with treatment effects on two surrogate endpoints}
\label{cross}
\end{table}

\section{Discussion}

We investigated the use of multiple surrogate endpoints as joint predictors of the clinical benefit measured on the final clinical outcome in advanced colorectal cancer.
A multivariate meta-analytic framework allowed us to combine treatment effects on two candidate surrogate endpoints (TR and PFS) and the treatment effect on the final clinical outcome (OS). In a Bayesian meta-analytic framework, we modelled the correlated treatment effect in the product normal formulation which is a convenient form to explore a range of parameters describing the surrogacy relationships, such as the intercept, slope, conditional variance (as set out by Daniels and Hughes \cite{daniels1997meta}) and the adjusted R-squared (introduced by Burzykowski et al \cite{burzykowski2001validation} and in the Bayesian framework by Renfro et al \cite{renfro2012bayesian}). These models also are used to make predictions of the treatment effect on the final clinical outcome (OS) from the treatment effect on the surrogate endpoints. In this respect they have an advantage of taking into account of the uncertainty around all the parameters, including the measurement error around the treatment effects on surrogate endpoints (in contrast to, for example, the standard approach to meta-regression where treatment effects on surrogate endpoints are treated as fixed covariates) \cite{bujkiewicz2015bayesian}.

The treatment effects on PFS and TR were associated with treatment effect on OS. However, overall the joint use of two surrogate endpoints did not lead to much improvement in the association between treatment effects on the surrogate and final endpoints but in the subclass of anti-angiogenic agents led to very modest improvement in precision of the predicted effects on OS. Some small improvement in precision, when modelling both surrogate endpoints jointly, was also observed in cross-validation procedure conducted on trials without cross-over.
In the trials allowing for cross-over, there is typically reduced effect on OS with large uncertainty around the treatment effect estimate. This is likely to affect  the results of modelling surrogate relationships, using both the bivariate and the trivariate methods. 
It is possible that the trivariate approach would show some noticeable benefit if more studies were available in the analysis of studies without the cross-over. In our analysis of the trials which did not allow for the cross-over, there was typically reduced uncertainty of the predicted effects when using multiple surrogates, but the reduction was small as the number of studies in the analysis was also small, therefore we cannot draw strong conclusions based on our findings. Not all studies reported whether the treatment cross-over was allowed. Another source of uncertainty that may have prevented the improvement of the surrogate relationship when using both candidate surrogate endpoints was the large between-studies heterogeneity of the treatment effect on the tumour response. This may have been caused by the heterogeneity of the methods used to measure the response across the trials.\\

To investigate sensitivity of the results to the model parameterisation, we carried out a sensitivity analysis, (the results of which are shown in  Table 6 of the online appendix). Using alternative assumptions about the between-studies variance-covariance structure gave similar results for the surroagte relationship between the treatment effects on PFS and OS, however the $R^{2}_{adjusted}$ was reduced compared to the bivariate and trivariate models considered in the main analysis, and for one of the models the interval of the slope included zero suggesting marginally poor surrogate relationship. \\ 

In conclusion, the impact of the joint modelling of the treatment effects on two surrogate endpoints (TR and PFS), on their surrogate relationship with the treatment effect on the overall survival was not noticeable in advance colorectal cancer. Further work will be needed to investigate in detail the treatment cross-over and the heterogeneity of the definitions of the outcomes and potentially the patient populations.

\section{Acknowledgements}
This work was supported by the Medical Research Council (MRC) Methodology
Research Programme [New Investigator Research Grant MR/L009854/1 awarded to Sylwia Bujkiewicz].
The research presented here used the ALICE High Performance Computing Facility at the University of Leicester.

\FloatBarrier
\newpage
\section*{References}


\newpage
\FloatBarrier

\appendix

	\begin{frontmatter}
	\title{Combining tumour response and progression free survival did not improve their value as surrogate endpoints for overall survival in advanced colorectal cancer: Supplementary material} 
	
	
	\author{Elia EG$^{1,2*}$, St{\"a}dler N$^{3}$, Ciani O$^{4,5}$, Taylor RS$^{4}$, Bujkiewicz S$^{1}$}  
	\fntext[myfootnote]{Department of Health Sciences, University of Leicester, George Davies Centre, University Road, Leicester, LE1 7RH, UK}
	
	\fntext[myfootnote]{Department of Biostatistics, Harvard University, 677 Huntington Ave, Boston, MA 02115, USA}
	
	\fntext[myfootnote]{F. Hoffmann-La Roche Ltd, Basel, Switzerland }
	
	\fntext[myfootnote]{Evidence Synthesis \& Modelling for Health Improvement, Institute of Health Research, University of Exeter Medical School, University of Exeter, Exeter EX2 4SG, UK }
	
	\fntext[myfootnote]{ CERGAS Bocconi University, via Rontgen 1, 20136 Milan, Italy}

	\cortext[correspondingauthor]{Corresponding author}
	\emailauthor{eelia@hsph.harvard.edu}{Elia EG}

\end{frontmatter}

\section{WinBUGS codes for bivariate and trivariate models}

\subsection{Bivariate model for PFS and OS}
\begin{verbatim}

model{
for (i in 1:num) {
rho_w[i] <-0.513  
prec_w[i,1:2,1:2] <- inverse(delta[i,1:2,1:2])	
#covariance matrix for the j-th study
delta[i,1,1]<-var[i,1]
delta[i,2,2]<-var[i,2] 
delta[i,1,2]<-sqrt(delta[i,1,1])*sqrt(delta[i,2,2])*rho_w[i]
delta[i,2,1]<-sqrt(delta[i,1,1])*sqrt(delta[i,2,2])*rho_w[i]
}

#Random effects model

for (i in 1:num) {
Y[i,1:2]~dmnorm(mu[i,1:2], prec_w[i,1:2,1:2])
#  product normal formulation for the between study part:
mu[i,1]~dnorm(rel,prec_rel)
mu[i,2]~dnorm(edss[i],prec_dis)
edss[i]<-lambda0+lambda1*mu[i,1]
}

rel~dnorm(0.0, 0.001)
prec_rel<-1/psi1.sq
prec_dis<-1/psi2.sq

lambda0~dnorm(0.0, 1.0E-3)
# prior between study correlations:
bcorrOSpfs~dunif(0,1)   
bcorr2<-pow(bcorrOSpfs,2)

sd.log.HR.pfs~dnorm(0,0.001)I(0,)
sd.log.HR.os~dnorm(0,0.001)I(0,)
psi1.sq<-sd.log.HR.pfs*sd.log.HR.pfs

sd.2.pfs<-psi1.sq 
sd.2.os<-sd.log.HR.os*sd.log.HR.os

#implied prior for lambda coefficients
lambda1<-bcorrOSpfs*sd.log.HR.os/sd.log.HR.pfs
psi2.sq<-pow(sd.log.HR.os,2) - pow(lambda1,2)*pow(sd.log.HR.pfs,2)



# estimates:
mean.log.HRpfs<-rel  
mean.log.HR.os<-lambda0+lambda1*rel  
mean.HR.pfs<-exp(mean.log.HRpfs) 
mean.HR.os<-exp(mean.log.HR.os) 
rhr.meanpfs.by.meanhros<-mean.HR.pfs/mean.HR.os
}


\end{verbatim}

\subsection{Trivariate model model for PFS, TR and OS}

\begin{verbatim}
Model { 
for (i in 1:num) {

rho_w_DM[i]<- -0.333  # response os
rho_w_RM[i]<- -0.433  # response pfs
rho_w_RD[i]<- 0.513   # pfs os 

Prec_w[i,1:3,1:3] <- inverse(delta[i,1:3,1:3])	
#covariance matrix for the j-th study
delta[i,1,1]<-var[i,1] 
delta[i,2,2]<-var[i,2] 
delta[i,3,3]<-var[i,3] 
delta[i,1,2]<-sqrt(delta[i,1,1])*sqrt(delta[i,2,2])*rho_w_RM[i]  
delta[i,2,1]<-delta[i,1,2]
delta[i,1,3]<-sqrt(delta[i,1,1])*sqrt(delta[i,3,3])*rho_w_DM[i]   
delta[i,3,1]<-delta[i,1,3]
delta[i,2,3]<-sqrt(delta[i,2,2])*sqrt(delta[i,3,3])*rho_w_RD[i]   
delta[i,3,2]<-delta[i,2,3]
}

# Random effects model

for (i in 1:num) {
Y[i,1:3]~dmnorm(mu[i,1:3], Prec_w[i,1:3,1:3])
#  product normal formulation for the between study part:
mu[i,1]~dnorm(etaM,precM)
mu[i,2]~dnorm(etaR[i],precR)
etaR[i]<-lambda20+lambda21*(mu[i,1]) 
mu[i,3]~dnorm(etaD[i],precD)
etaD[i]<-lambda30+lambda23*(mu[i,2])  
}

etaM~dnorm(0.0, 0.001)
lambda30~dnorm(0.0, 1.0E-3)  
lambda20~dnorm(0.0, 1.0E-3)   


bcorr12~dunif(-1,0)  
bcorr32~dunif(0,1)  

bcorr31<-bcorr12*bcorr32
bcorr31.sq<-pow(bcorr31,2)
bcorr12.sq<-pow(bcorr12,2)
bcorr32.sq<-pow(bcorr32,2)

sd.res~dnorm(0,0.001)I(0,)
sd.dis~dnorm(0,0.001)I(0,)
sd.pfs~dnorm(0,0.001)I(0,)
tau1.sq<-pow(sd.res,2)
tau3.sq<-pow(sd.os,2)
tau2.sq<-pow(sd.pfs,2)



psi1.sq<-pow(sd.res,2)    
precM<-1/psi1.sq

psi2.sq<-pow(sd.pfs,2) - pow(lambda21,2)*pow(sd.res,2)  
precR<-1/psi2.sq

psi3.sq<-pow(sd.os,2) - pow(lambda23,2)*pow(sd.pfs,2)   
precD<-1/psi3.sq


mean.res<-etaM #mean log OR response
mean.pfs<-lambda20+lambda21*etaM  #mean log hr pfs 
mean.os<-lambda30+lambda23*mean.pfs

lambda21<-bcorr12*sd.pfs/sd.res   
lambda23<-(bcorr32*sd.os/sd.pfs )   

meanORresp<-exp(mean.res)            
meanHRpfs<-exp(mean.pfs)               
meanHRos<-exp(mean.os)	

rhr.meanpfs.by.meanhros<-meanHRpfs/meanHRos  
}

\end{verbatim}

\section{Summary of the data}

Figure 1 shows forest plots for the treatment effect on the three outcomes: TR, PFS and OS. on log OR, log HR and logHR scales respectively. Figures 2-4 include the scatter plots showing the association patterns between the treatment effects on each pair of the three outcomes.

\begin{landscape}
	
	\begin{figure}[h]
		\caption{Forest plots of effect size estimate for each study for the complete dataset for outcomes: Response (Hazard ratio estimate), PFS (Odds ratio estimate), OS (odds ratio estimate) }
		\includegraphics[width=25cm]{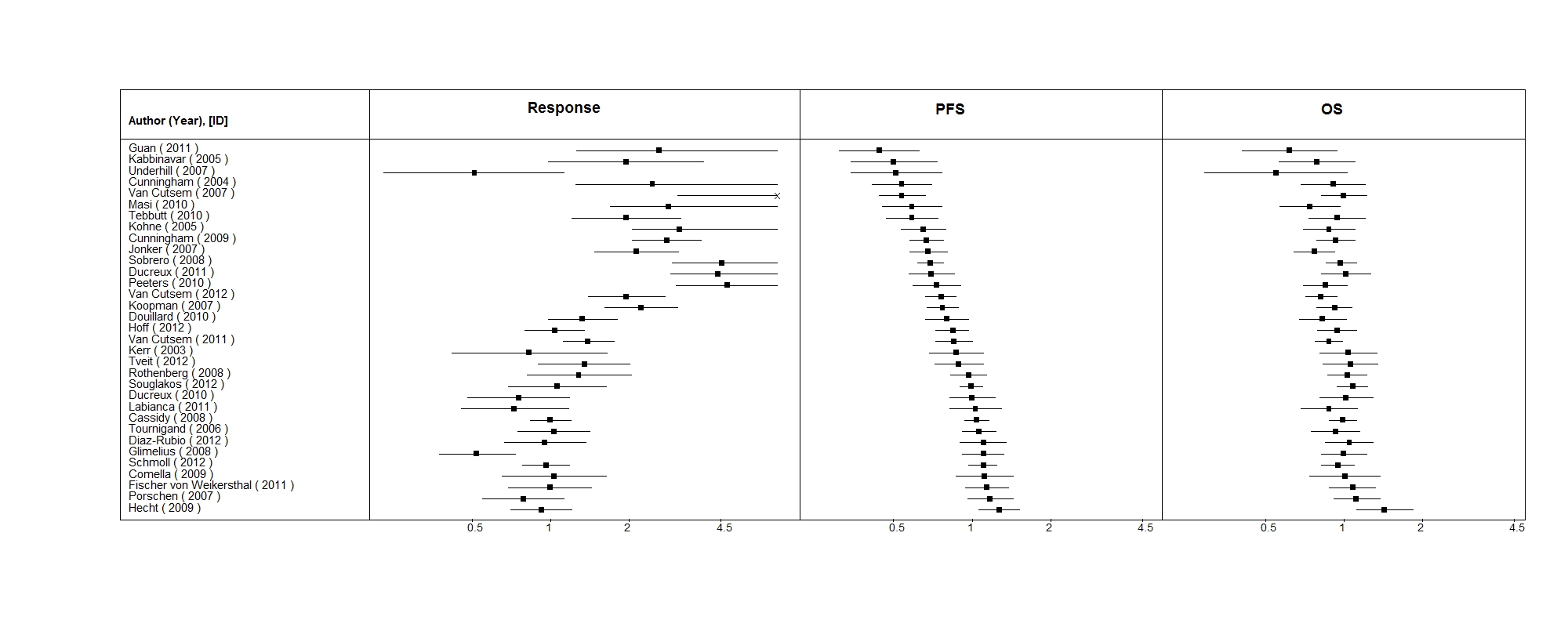}
		\label{fig:allforest}
	\end{figure}
\end{landscape}

\begin{figure}[!ht]
	\centering
	\caption{Scatter plot of effect size estimate for each subclass therapy for the base case scenario for outcomes: PFS (log hazard ratio), OS (log hazard ratio) }
	
	\includegraphics[width=0.49\linewidth]{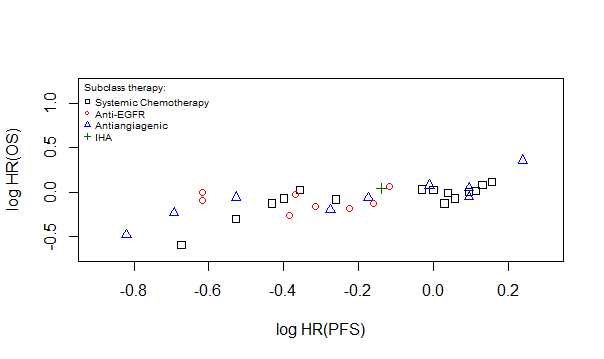}
	\label{fig:PFSOSscatter}
\end{figure}%

\begin{figure} [!ht]
	\centering
	\caption{Scatter plot of effect size estimate for each subclass therapy for the base case scenario for outcomes: Tumour Response (log odds ratio ), OS (log hazard ratio) }
	\includegraphics[width=.5\linewidth]{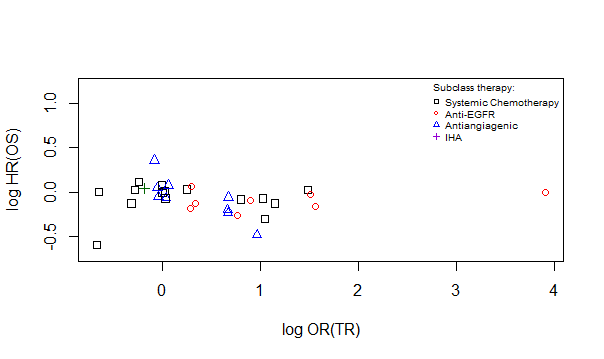}
	\label{fig:TROSscatter}
\end{figure}

\begin{figure} [!ht]
	\centering
	\caption{Scatter plot of effect size estimate for each subclass therapy for the base case scenario for outcomes: Tumour response (log odds ratio), PFS (log hazard ratio)
	}
	\includegraphics[width=.5\linewidth]{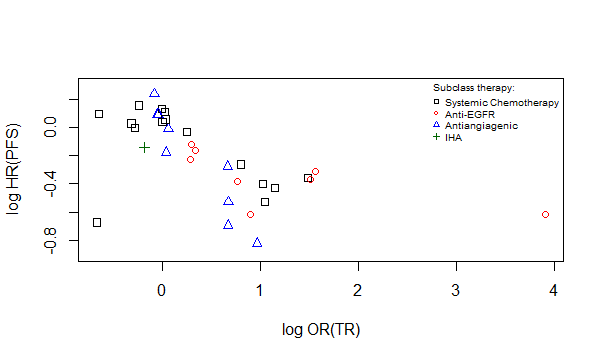}
	\label{fig:TRPFSscatter}
\end{figure}

\section{Additional results from base case scenario}

\begin{table}[h!]
	\centering
	\scalebox{0.8}{	\begin{tabular}{cccccccc}
			

			&1D TR&	1D PFS&1D OS&2D TR PFS& 2D TR OS&2D PFS OS&3D\\

			\hline
			
			\multicolumn{8}{c}{\emph{All treatments }} \\

			mean(OR(TR))&1.49(1.18,1.85)&&&1.52(1.21,1.88)&1.46(1.17,1.81)&&1.52(1.21,1.88)\\   
			mean(HR(PFS))&&0.82(0.74,0.9)&&0.83(0.76,0.91)&&0.82(0.75,0.9)&0.83(0.76,0.91)\\    
			mean(HR(OS))&&&0.95(0.91,0.99)&&0.95(0.91,0.99)&0.94(0.9,0.99)&0.95(0.91,0.99)\\    
			$\tau^{2}_{TR}$&0.36(0.19,0.65)&&&0.36(0.19,0.63)&0.36(0.19,0.64)&&0.36(0.19,0.65)\\ 
			$\tau^{2}_{PFS}$&&0.07(0.04,0.13)&&0.06(0.03,0.11)&&0.06(0.03,0.11)&0.05(0.03,0.1)\\ 
			$\tau^{2}_{OS}$&&&0(0,0.01)&&0(0,0.02)&0.01(0,0.02)&0.01(0,0.02)\\

			\multicolumn{8}{c}{\emph{Systemic chemotherapy }} \\

			mean(OR(TR))&1.31(0.87,1.89)&&&1.32(0.88,1.9)&1.29(0.84,1.88)&&1.3(0.87,1.86)\\     
			mean(HR(PFS))&&0.89(0.76,1.02)&&0.9(0.78,1.02)&&0.89(0.77,1.02)&0.9(0.79,1.02)\\    
			mean(HR(OS))&&&0.97(0.91,1.02)&&0.97(0.91,1.03)&0.96(0.9,1.03)&0.97(0.91,1.03)\\    
			$\tau^{2}_{TR}$&0.53(0.21,1.23)&&&0.5(0.2,1.16)&0.56(0.22,1.3)&&0.51(0.2,1.16)\\     
			$\tau^{2}_{PFS}$&&0.07(0.02,0.17)&&0.06(0.02,0.15)&&0.06(0.02,0.14)&0.05(0.02,0.12)\\
			$\tau^{2}_{OS}$&&&0(0,0.01)&&0(0,0.01)&0(0,0.02)&0(0,0.01)\\

			\multicolumn{8}{c}{\emph{anti-EGFR  therapies   }} \\

			mean(OR(TR))&2.57(1.34,4.88)&&&2.66(1.36,5.12)&2.64(1.28,5.45)&&2.81(1.33,5.75)\\   
			mean(HR(PFS))&&0.71(0.6,0.83)&&0.72(0.62,0.83)&&0.71(0.58,0.85)&0.72(0.6,0.85)\\    
			mean(HR(OS))&&&0.9(0.82,0.99)&&0.91(0.82,1.01)&0.9(0.82,0.99)&0.91(0.82,1.01)\\     
			$\tau^{2}_{TR}$&0.71(0.11,2.91)&&&0.73(0.12,2.94)&0.91(0.13,3.87)&&0.96(0.14,3.92)\\ 
			$\tau^{2}_{PFS}$&&0.04(0,0.17)&&0.03(0,0.13)&&0.06(0.01,0.23)&0.05(0.01,0.18)\\      
			$\tau^{2}_{OS}$&&&0.01(0,0.04)&&0.01(0,0.06)&0.01(0,0.06)&0.01(0,0.06)\\    
			
			\multicolumn{8}{c}{\emph{Anti-angiogenic agents   }} \\	 
			mean(OR(TR))&1.31(0.96,1.8)&&&1.34(0.98,1.86)&1.3(0.96,1.78)&&1.33(1,1.82)\\        
			mean(HR(PFS))&&0.82(0.6,1.08)&&0.82(0.61,1.07)&&0.82(0.61,1.06)&0.83(0.63,1.06)\\   
			mean(HR(OS))&&&0.96(0.81,1.12)&&0.96(0.82,1.1)&0.95(0.8,1.1)&0.95(0.81,1.1)\\       
			$\tau^{2}_{TR}$&0.17(0.02,0.62)&&&0.19(0.03,0.63)&0.16(0.02,0.57)&&0.16(0.02,0.52)\\ 
			$\tau^{2}_{PFS}$&&0.19(0.04,0.6)&&0.16(0.04,0.49)&&0.16(0.04,0.49)&0.14(0.03,0.4)\\  
			$\tau^{2}_{OS}$&&&0.04(0,0.16)&&0.04(0,0.15)&0.04(0,0.17)&0.04(0,0.13)\\

			\hline	
	\end{tabular}}
	\caption{Heterogeneity and summary measures and associated 95\% Credible Intervals obtained from the bivariate and structured trivariate models, applied to the complete data;  surrogate outcomes TR: Treatment response rate, PFS: Progression free survival and final outcome OS: Overall survival, 1D:Univariate model Posterior mean (95\% CrI), 2D: Bivariate model Posterior mean (95\% CrI), 3D: Trivariate model Posterior mean (95\% CrI)}
	\label{appbivcom}
\end{table}

\begin{table}[ht]
	\centering
	
	\scalebox{0.7}{	\begin{tabular}{ccccc}
			
			Study index&	Observed OS	& bivariate &  trivariate structured&\%red \\
			\hline
			1&0.99(0.88,1.12)&1(0.84,1.18)&0.99(0.83,1.18)&0.04\\         2&1.01(0.74,1.38)&1(0.71,1.4)&0.99(0.71,1.38)&1.02\\       								
			3&0.91(0.68,1.21)&0.87(0.63,1.2)&0.89(0.65,1.22)&-1.48\\      4&0.93(0.79,1.11)&0.9(0.73,1.11)&0.91(0.74,1.12)&-0.38\\   								
			5&1.05(0.85,1.3)&1(0.78,1.28)&0.99(0.77,1.27)&1.34\\          6&0.83(0.67,1.02)&0.94(0.74,1.2)&0.95(0.75,1.2)&-0.08\\    								
			7&1.02(0.8,1.3)&0.98(0.75,1.28)&0.98(0.75,1.28)&0.12\\        8&1.02(0.82,1.27)&0.91(0.71,1.17)&0.91(0.71,1.17)&-0.39\\  								
			9&1.08(0.88,1.32)&1(0.79,1.28)&1(0.78,1.27)&0.71\\            10&1(0.82,1.22)&1(0.79,1.27)&1(0.79,1.27)&-0.66\\           								
			11&0.62(0.41,0.94)&0.87(0.56,1.36)&0.89(0.57,1.39)&-1.58\\     12&1.43(1.12,1.84)&1.01(0.76,1.32)&1(0.76,1.31)&1.54\\      								
			13&0.94(0.79,1.12)&0.95(0.77,1.17)&0.96(0.77,1.18)&-0.62\\     14&0.77(0.64,0.92)&0.92(0.74,1.13)&0.93(0.75,1.14)&-0.66\\  								
			15&0.79(0.57,1.11)&0.88(0.61,1.27)&0.91(0.63,1.3)&-1.53\\      16&1.04(0.81,1.34)&0.95(0.72,1.26)&0.96(0.72,1.27)&-0.35\\  								
			17&0.88(0.7,1.11)&0.9(0.69,1.17)&0.91(0.7,1.18)&-0.73\\        18&0.92(0.79,1.07)&0.93(0.76,1.13)&0.94(0.77,1.14)&0.41\\   								
			19&0.88(0.68,1.13)&0.99(0.75,1.31)&0.99(0.75,1.31)&-0.65\\     20&0.74(0.57,0.97)&0.9(0.67,1.21)&0.91(0.68,1.22)&-0.49\\   								
			21&0.85(0.7,1.04)&0.93(0.73,1.17)&0.93(0.73,1.16)&0.29\\       22&1.12(0.91,1.38)&1.01(0.79,1.28)&1(0.79,1.26)&1.96\\      								
			23&1.03(0.87,1.22)&0.97(0.79,1.2)&0.97(0.79,1.2)&0\\           24&0.95(0.82,1.11)&1.02(0.84,1.23)&1.01(0.83,1.22)&0.43\\   								
			25&0.97(0.85,1.12)&0.9(0.76,1.08)&0.91(0.76,1.08)&-1.63\\      26&1.08(0.94,1.25)&0.98(0.82,1.16)&0.97(0.82,1.16)&0.69\\   								
			27&0.94(0.73,1.21)&0.88(0.66,1.18)&0.9(0.68,1.2)&-0.88\\       28&0.93(0.75,1.15)&1(0.78,1.28)&1(0.78,1.27)&0.49\\         								
			29&1.06(0.84,1.35)&0.96(0.73,1.25)&0.96(0.73,1.25)&0.16\\      30&0.55(0.29,1.04)&0.89(0.46,1.7)&0.93(0.49,1.78)&-4.4\\    								
			31&1(0.82,1.22)&0.85(0.66,1.08)&0.87(0.68,1.1)&-0.36\\         32&0.88(0.77,1)&0.96(0.81,1.13)&0.96(0.81,1.13)&0.86\\      								
			33&0.82(0.71,0.94)&0.94(0.79,1.11)&0.94(0.79,1.12)&-0.37\\

			\multicolumn{3}{l}{Average (range) \% reduction} & \multicolumn{2}{r}{-0.21 (-4.40, 1.96)}\\
			\hline
	\end{tabular} }
	\caption{Predictions of the treatment effect on OS obtained from the bivariate model (with PFS as surrogate endpoint) and the trivariate model (with TR and PFS as two surrogate endpoints modelled jointly) presented alongside of the observed estimates for the base case scenario (33 studies); \% red refers to the percentage reduction in the width of the predicted interval when comparing the predicted effects obtained from the trivariate model with the effects predicted using the bivariate model. The study index corresponds to the number in the reference list for the base case scenario included in the supplementary materials.}
	\label{preco}
\end{table}

\begin{table}[ht]
	\centering
	\scalebox{0.7}{		\begin{tabular}{rrrrrrrr }

			Subclass&	Study index*&	Observed OS	& bivariate &  trivariate structured&\%red& \\
			\hline

			Systemic chemotherapy	&1&0.99(0.88,1.12)&0.99(0.84,1.17)&0.99(0.84,1.16)&4.16\\    &2&1.01(0.74,1.38)&1(0.71,1.39)&0.99(0.71,1.37)&1.85\\      								
			&4&0.93(0.79,1.11)&0.92(0.73,1.14)&0.93(0.74,1.15)&1.48\\    &7&1.02(0.8,1.3)&0.98(0.75,1.27)&0.98(0.76,1.27)&1.01\\     								
			&8&1.02(0.82,1.27)&0.92(0.71,1.18)&0.92(0.71,1.18)&-0.13\\   &9&1.08(0.88,1.32)&1(0.78,1.26)&0.99(0.78,1.24)&3.22\\      								
			&10&1(0.82,1.22)&1(0.79,1.26)&1(0.79,1.25)&0.83\\               &17&0.88(0.7,1.11)&0.93(0.71,1.21)&0.93(0.72,1.21)&1.7\\     							
			&18&0.92(0.79,1.07)&0.95(0.78,1.15)&0.95(0.78,1.15)&1.07\\      &19&0.88(0.68,1.13)&0.99(0.75,1.3)&0.99(0.76,1.3)&0.39\\     							
			&20&0.74(0.57,0.97)&0.94(0.69,1.26)&0.94(0.7,1.26)&1.08\\       &22&1.12(0.91,1.38)&0.99(0.79,1.25)&0.99(0.78,1.24)&3.14\\   							
			&23&1.03(0.87,1.22)&0.97(0.79,1.19)&0.97(0.79,1.19)&1.79\\      &28&0.93(0.75,1.15)&1(0.78,1.27)&0.99(0.78,1.26)&2.42\\      							&30&0.55(0.29,1.04)&0.92(0.48,1.76)&0.96(0.5,1.82)&-2.92\\  
			Average \% reduction (range)&&&&&1.41 (-2.92,4.16)\\ 								
			
			Anti-EGFR	&3&0.91(0.68,1.21)&0.88(0.59,1.28)&0.89(0.59,1.31)&-4.82\\      &6&0.83(0.67,1.02)&0.94(0.67,1.28)&0.95(0.67,1.32)&-6.72\\   							&14&0.77(0.64,0.92)&0.92(0.7,1.2)&0.93(0.7,1.22)&-6.72\\       &21&0.85(0.7,1.04)&0.92(0.67,1.25)&0.92(0.65,1.28)&-6.81\\    							&25&0.97(0.85,1.12)&0.89(0.68,1.15)&0.89(0.66,1.18)&-10.33\\     &29&1.06(0.84,1.35)&0.92(0.65,1.27)&0.93(0.64,1.32)&-9.54\\   						&31&1(0.82,1.22)&0.85(0.62,1.15)&0.85(0.62,1.16)&-0.77\\         &32&0.88(0.77,1)&0.95(0.7,1.27)&0.96(0.69,1.32)&-9.58\\      
			Average \% reduction (range)&&&&&-6.91 (-9.58,-0.77)\\ 	 								
			
			Antiangiogenic	&5&1.05(0.85,1.3)&1.07(0.71,1.58)&1.07(0.71,1.56)&3.3\\        &11&0.62(0.41,0.94)&0.83(0.47,1.43)&0.84(0.48,1.44)&-0.07\\  							
			&12&1.43(1.12,1.84)&1.04(0.72,1.47)&1.03(0.72,1.44)&4.55\\      &13&0.94(0.79,1.12)&0.97(0.66,1.38)&0.97(0.67,1.37)&3.34\\   							
			&15&0.79(0.57,1.11)&0.82(0.49,1.35)&0.84(0.5,1.36)&-0.38\\      &24&0.95(0.82,1.11)&1.11(0.78,1.55)&1.11(0.78,1.53)&2.46\\   							
			&26&1.08(0.94,1.25)&1.02(0.71,1.41)&1.01(0.72,1.39)&4.4\\       &27&0.94(0.73,1.21)&0.81(0.53,1.23)&0.83(0.54,1.23)&1.67\\  
			&	33&0.82(0.71,0.94)&0.96(0.7,1.29)&0.96(0.7,1.29)&1.91\\
			Average \% reduction (range)&&&&&2.35 (-0.07,4.4)\\

			\hline
			
	\end{tabular} }
	\caption{{\scriptsize Predicted treatment effects on OS obtained in the cross validation procedure in subgroup analyses (class of therapy) from the bivariate model (with PFS as surrogate endpoint) and the trivariate model (with TR and PFS as two surrogate endpoints modelled jointly) presented alongside of the observed estimates for the base case scenario (33 studies); \% red refers to the percentage reduction in the width of the predicted interval corresponding to the prediction from the trivariate model compared to the bivariate model}. The study index corresponds to the number in the reference list for the base case scenario included in the supplementary materials.}
	\label{predsub}	
\end{table}

\newpage

\section{Additional results from sensitivity analyses}
\subsection{Set of 51 trials reporting at least two outcomes}

\begin{table}[h!]
	\centering
	\begin{tabular}{cccccc}
		&Response OS&Response PFS&Response PFS   &PFS OS&PFS OS                                       \\
		&2D&2D& 3D & 2D&3D\\

		\hline
		\multicolumn{6}{c}{\emph{All treatments}} \\
		
		intercept&-0.01(-0.05,0.04)&-0.05(-0.14,0.03)&-0.04(-0.12,0.03)&0(-0.05,0.05)&0.01(-0.03,0.05)\\     
		slope&-0.09(-0.17,-0.02)&-0.33(-0.46,-0.19)&-0.33(-0.46,-0.21)&0.23(0.04,0.43)&0.27(0.1,0.44)\\      
		variance&0(0,0.01)&0.03(0.01,0.05)&0.02(0.01,0.04)&0(0,0.01)&0(0,0.01)\\                             
		$R^{2}_{adjusted}$&0.39(0.02,0.85)&0.57(0.24,0.83)&0.59(0.28,0.83)&0.57(0.06,0.96)&0.57(0.13,0.94)\\   
		
		\\
		\multicolumn{6}{c}{\emph{Systemic chemotherapy }} \\

		intercept&-0.01(-0.06,0.04)&-0.06(-0.17,0.03)&-0.04(-0.14,0.04)&-0.02(-0.08,0.04)&0(-0.05,0.05)\\    
		slope&-0.08(-0.17,-0.01)&-0.25(-0.41,-0.08)&-0.29(-0.43,-0.13)&0.17(0.01,0.45)&0.27(0.05,0.5)\\      
		variance&0(0,0.01)&0.02(0,0.07)&0.02(0,0.05)&0(0,0.01)&0(0,0.01)\\                                   
		$R^{2}_{adjusted}$&0.5(0.01,0.97)&0.54(0.06,0.94)&0.62(0.16,0.96)&0.53(0.01,0.98)&0.61(0.04,0.98)\\    
		\\
		\multicolumn{6}{c}{\emph{Anti-EGFR  therapies }} \\
		
		intercept&-0.05(-0.14,0.07)&-0.21(-0.37,0.01)&-0.15(-0.33,0.07)&-0.04(-0.15,0.19)&-0.02(-0.12,0.18)\\
		slope&-0.04(-0.17,0)&-0.14(-0.36,-0.01)&-0.18(-0.41,-0.02)&0.16(0,0.74)&0.17(0,0.74)\\               
		variance&0.01(0,0.04)&0.02(0,0.08)&0.02(0,0.1)&0.01(0,0.05)&0.01(0,0.04)\\                           
		$R^{2}_{adjusted}$&0.21(0,0.83)&0.45(0,0.96)&0.49(0.01,0.96)&0.23(0,0.86)&0.21(0,0.84)\\      
		
		\\
		\multicolumn{6}{c}{\emph{Anti-angiogenic agents }} \\

		intercept&0.09(-0.07,0.33)&0.07(-0.14,0.29)&0.07(-0.13,0.29)&0.04(-0.06,0.14)&0.06(-0.04,0.17)\\     
		slope&-0.47(-1.24,-0.03)&-0.94(-1.82,-0.32)&-0.9(-1.86,-0.24)&0.37(0.05,0.75)&0.41(0.04,0.85)\\      
		variance&0.04(0,0.16)&0.04(0,0.14)&0.03(0,0.12)&0.01(0,0.04)&0.01(0,0.06)\\                          
		$R^{2}_{adjusted}$&0.43(0,0.94)&0.72(0.11,0.99)&0.64(0.06,0.98)&0.58(0.03,0.97)&0.54(0.02,0.96)\\
		
		\hline	
		
	\end{tabular}
	\caption{Surrogacy criteria: posterior means, between study variances and associated 95\% Credible Intervals obtained from the bivariate and structured trivariate models, applied to data where at least two outcomes are reported for each trial (51 trials); surrogate outcomes TR: Treatment response rate, PFS: Progression free survival and final outcome OS: Overall survival}
	\label{bivcom52}
\end{table}

\begin{table}[h!]
	\centering
	\scalebox{0.78}{		\begin{tabular}{ccccccccccc}
			
			&Response OS&Response OS& PFS OS&   &                                    \\
			&2D&2D&\ 2D&3D\\	
			\hline

			\multicolumn{6}{l}{\emph{All treatments}} \\
			mean(OR(TR))&1.48(1.2,1.81)&1.41(1.18,1.68)&&1.42(1.21,1.67)\\   
			mean(HR(PFS))&0.84(0.77,0.91)&&0.83(0.76,0.9)&0.85(0.79,0.92)\\  
			mean(HR(OS))&&0.96(0.92,1)&0.95(0.91,1)&0.97(0.93,1)\\           
			$\tau^{2}_{TR}$&0.32(0.18,0.56)&0.3(0.17,0.49)&&0.27(0.16,0.44)\\ 
			$\tau^{2}_{PFS}$&0.06(0.03,0.11)&&0.06(0.03,0.1)&0.05(0.03,0.09)\\
			$\tau^{2}_{OS}$&&0.01(0,0.02)&0.01(0,0.02)&0.01(0,0.02)\\

			\multicolumn{6}{l}{\emph{Systemic chemotherapy }} \\	 
			mean(OR(TR))&1.3(0.92,1.79)&1.29(0.97,1.68)&&1.3(1.01,1.65)\\    
			mean(HR(PFS))&0.88(0.78,1)&&0.89(0.77,1.02)&0.89(0.8,0.99)\\     
			mean(HR(OS))&&0.97(0.92,1.03)&0.96(0.9,1.02)&0.97(0.91,1.02)\\   
			$\tau^{2}_{TR}$&0.43(0.18,0.95)&0.39(0.19,0.76)&&0.34(0.17,0.64)\\
			$\tau^{2}_{PFS}$&0.05(0.02,0.12)&&0.06(0.02,0.14)&0.05(0.02,0.1)\\
			$\tau^{2}_{OS}$&&0.01(0,0.02)&0(0,0.02)&0.01(0,0.02)\\

			\\
			\multicolumn{6}{l}{\emph{Anti-EGFR  therapies }} \\
			
			mean(OR(TR))&2.66(1.36,5.12)&2.13(1.23,3.65)&&2.13(1.2,3.67)\\   
			mean(HR(PFS))&0.72(0.62,0.83)&&0.73(0.61,0.88)&0.76(0.65,0.9)\\  
			mean(HR(OS))&&0.93(0.85,1.01)&0.91(0.83,1.02)&0.93(0.86,1.03)\\  
			$\tau^{2}_{TR}$&0.73(0.12,2.94)&0.62(0.13,2.19)&&0.71(0.14,2.37)\\
			$\tau^{2}_{PFS}$&0.03(0,0.13)&&0.06(0.01,0.22)&0.05(0.01,0.17)\\  
			$\tau^{2}_{OS}$&&0.01(0,0.04)&0.01(0,0.08)&0.01(0,0.05)\\

			\multicolumn{6}{l}{\emph{Anti-angiogenic agents }} \\	
			mean(OR(TR))&1.31(1,1.75)&1.29(0.98,1.73)&&1.28(1.04,1.61)\\     
			mean(HR(PFS))&0.86(0.65,1.1)&&0.83(0.67,1.01)&0.88(0.72,1.05)\\  
			mean(HR(OS))&&1(0.83,1.21)&0.97(0.85,1.08)&1.01(0.89,1.15)\\     
			$\tau^{2}_{TR}$&0.15(0.02,0.5)&0.15(0.02,0.49)&&0.1(0.01,0.29)\\  
			$\tau^{2}_{PFS}$&0.16(0.04,0.45)&&0.11(0.03,0.3)&0.1(0.03,0.26)\\ 
			$\tau^{2}_{OS}$&&0.07(0,0.26)&0.03(0,0.1)&0.04(0,0.12)\\		
			
			\\
			\hline	
			
	\end{tabular} }
	\caption{{\small Posterior means and between study variances and associated 95\% Credible Intervals obtained from the bivariate and structured trivariate models for each subclass therapy, using data where at least two outcomes are reported for each trial (51 trials);  surrogate outcomes TR: Treatment response rate, PFS: Progression free survival and final outcome OS: Overall survival }}
	\label{class52}
\end{table}

\newpage

\FloatBarrier
\subsection{Set of 7 trials not allowing for crossover}

\begin{table}[h!]
	\centering
	\begin{tabular}{cccccccccc}	
		&2D TR PFS&2D TR OS &2D PFS OS &3D   \\
		\hline

		&mean(OR(TR))&1.32(0.8,2.02)&1.32(0.76,2.08)& &1.33(0.81,2.03)\\    
		&mean(HR(PFS))&0.88(0.73,1.06)& &0.88(0.72,1.07)&0.88(0.73,1.05)\\  
		&mean(HR(OS))& &0.94(0.83,1.06)&0.94(0.83,1.06)&0.93(0.83,1.06)\\   
		&$\tau^{2}_{TR}$&0.34(0.06,1.26)&0.41(0.07,1.58)& &0.33(0.06,1.23)\\ 
		&$\tau^{2}_{PFS}$&0.05(0.01,0.21)& & &0.05(0.01,0.2)\\       
		&$\tau^{2}_{OS}$& &0.01(0,0.07)& &0.01(0,0.06)\\

		\hline	
	\end{tabular}
	\caption{Mean effects and heterogeneity parameters obtained from the bivariate models and structured trivariate model applied to the trials that did not crossover therapy (7 studies complete data) }
	\label{crossapp}
\end{table}

\begin{table}[ht]
	\centering
	\scalebox{0.7}{	\begin{tabular}{rrrrrrr }
			\hline

			Study index*&	Observed OS	& bivariate &  trivariate structured&\%red& \\
			\hline
			
			2&1.01(0.74,1.38)&1(0.66,1.49)&0.99(0.65,1.46)&3.24\\     4&0.93(0.79,1.11)&0.84(0.59,1.16)&0.83(0.59,1.16)&-0.31\\
			5&1.05(0.85,1.3)&0.99(0.69,1.38)&0.97(0.69,1.35)&5.51\\   6&0.83(0.67,1.02)&0.94(0.68,1.28)&0.94(0.69,1.27)&1.77\\
			7&1.02(0.8,1.3)&0.97(0.68,1.35)&0.97(0.68,1.36)&-1.75\\   16&1.04(0.81,1.34)&0.93(0.66,1.3)&0.93(0.66,1.29)&1.63\\
			33&0.82(0.71,0.94)&0.94(0.72,1.22)&0.94(0.73,1.2)&6.19\\
			
			Average \% reduction (range)&&&&2.33(-1.74,6.19)\\
			\hline
	\end{tabular}}
	\caption{Predictions of the treatment effect on OS obtained from the bivariate model (with PFS as surrogate endpoint) and the trivariate model (with TR and PFS as two surrogate endpoints modelled jointly) presented alongside of the observed estimates  for 7 studies with no cross over; \% red refers to the percentage reduction in the width of the predicted interval when comparing the predicted effects obtained from the trivariate model with the effects predicted using the bivariate model. The study index corresponds to the number in the reference list for the base case scenario included in the supplementary materials.}
	
	\label{crosspred}
\end{table}

\newpage
\FloatBarrier
\subsection{Set of 32 studies, sensitivity analysis results where outlier 1 study was removed (index 93) }

\begin{table}[h!]
	\centering
	\begin{tabular}{cccccccccc}
		&Response OS&Response PFS&Response PFS   &PFS OS&PFS OS                                       \\
		&2D&2D& 3D & 2D&3D\\

		\hline
		\multicolumn{6}{l}{\emph{All treatments}} \\
		intercept&-0.03(-0.08,0.02)&-0.05(-0.14,0.03)&-0.05(-0.14,0.02)&-0.01(-0.06,0.04)&-0.01(-0.06,0.03)\\
		slope&-0.06(-0.13,0)&-0.32(-0.45,-0.19)&-0.31(-0.43,-0.18)&0.28(0.08,0.48)&0.26(0.06,0.46)\\         
		variance&0(0,0.01)&0.02(0.01,0.05)&0.02(0.01,0.04)&0(0,0.01)&0(0,0.01)\\                           
		$R^{2}_{adjusted}$&0.36(0,0.92)&0.6(0.24,0.86)&0.61(0.26,0.88)&0.68(0.15,0.98)&0.63(0.1,0.98)\\
		\\

		\multicolumn{6}{l}{\emph{Systemic chemotherapy }} \\		intercept&-0.02(-0.08,0.04)&-0.04(-0.16,0.06)&-0.04(-0.14,0.05)&-0.02(-0.08,0.04)&-0.02(-0.08,0.04)\\ 
		slope&-0.03(-0.11,0)&-0.26(-0.42,-0.08)&-0.25(-0.4,-0.09)&0.17(0,0.45)&0.14(0,0.4)\\                  
		variance&0(0,0.01)&0.02(0,0.08)&0.02(0,0.06)&0(0,0.01)&0(0,0.01)\\                                    
		$R^{2}_{adjusted}$&0.39(0,0.96)&0.58(0.07,0.96)&0.66(0.11,0.98)&0.52(0.01,0.98)&0.47(0,0.97)\\        
		\\
		\multicolumn{6}{l}{\emph{Anti-EGFR  therapies }} \\
		intercept&-0.08(-0.18,0.09)&-0.21(-0.36,0.02)&-0.19(-0.36,0.05)&0(-0.17,0.61)&0.03(-0.17,0.86)\\      
		slope&-0.05(-0.21,0)&-0.12(-0.37,0)&-0.14(-0.4,-0.01)&0.38(0,2.3)&0.49(0,3.12)\\                 variance&0.01(0,0.06)&0.02(0,0.08)&0.02(0,0.1)&0.01(0,0.05)&0.01(0,0.05)\\                        $R^{2}_{adjusted}$&0.23(0,0.87)&0.39(0,0.94)&0.4(0,0.95)&0.28(0,0.9)&0.27(0,0.89)\\                   
		\\
		\multicolumn{6}{l}{\emph{Anti-angiogenic agents }} \\  intercept&0.04(-0.09,0.2)&0.03(-0.18,0.25)&0.03(-0.18,0.25)&0.02(-0.1,0.15)&0.02(-0.09,0.14)\\        
		slope&-0.35(-0.89,-0.03)&-0.87(-1.64,-0.3)&-0.85(-1.64,-0.27)&0.38(0.05,0.79)&0.37(0.04,0.77)\\       
		variance&0.02(0,0.07)&0.04(0,0.15)&0.03(0,0.14)&0.02(0,0.06)&0.01(0,0.06)\\                     $R^{2}_{adjusted}$&0.52(0.01,0.97)&0.74(0.13,0.99)&0.72(0.1,0.99)&0.59(0.03,0.97)&0.56(0.02,0.97)\\   
		\hline 
		
	\end{tabular}
	
	\caption{Surrogacy criteri: posterior means, between study variances and associated 95\% Credible Intervals obtained from the bivariate and structured trivariate models, applied to data (33 trials), removing outlier trial indexed 93;  surrogate outcomes TR: Treatment response rate, PFS: Progression free survival and final outcome OS: Overall survival}
	\label{sensrm93}
\end{table}

\subsection{    Set of 33 studies, sensitivity analysis results where the prior for between-study correlation is set to U(-1,1)}

\begin{table} 
	\centering
	\begin{tabular}{cccccc}
		&Response OS&Response PFS&Response PFS   &PFS OS&PFS OS                                       \\
		&2D&2D& 3D & 2D&3D\\	
		\hline
		\multicolumn{6}{c}{\emph{All treatments}} \\

		intercept&-0.03(-0.08,0.02)&-0.05(-0.14,0.03)&-0.05(-0.13,0.02)&-0.01(-0.06,0.03)&-0.02(-0.07,0.03)\\
		slope&-0.04(-0.13,0.02)&-0.32(-0.45,-0.2)&-0.31(-0.43,-0.19)&0.22(0.02,0.41)&0.19(0,0.4)\\           
		variance&0(0,0.01)&0.02(0.01,0.05)&0.02(0.01,0.04)&0(0,0.01)&0(0,0.01)\\                             
		$R^{2}_{adjusted}$&0.31(0,0.9)&0.61(0.27,0.87)&0.64(0.3,0.89)&0.58(0.05,0.97)&0.49(0.01,0.95)\\      
		
		\multicolumn{6}{c}{\emph{Systemic chemotherapy }} \\		
		
		intercept&-0.03(-0.09,0.03)&-0.05(-0.16,0.05)&-0.04(-0.14,0.05)&-0.02(-0.08,0.04)&-0.02(-0.08,0.04)\\
		slope&-0.02(-0.1,0.04)&-0.26(-0.42,-0.07)&-0.25(-0.4,-0.09)&0.14(-0.07,0.44)&0.1(-0.1,0.39)\\        
		variance&0(0,0.01)&0.03(0,0.08)&0.02(0,0.06)&0(0,0.01)&0(0,0.01)\\                                   $R^{2}_{adjusted}$&0.35(0,0.96)&0.58(0.05,0.96)&0.66(0.11,0.98)&0.49(0,0.98)&0.43(0,0.97)\\          
		
		\multicolumn{6}{c}{\emph{Anti-EGFR  therapies }} \\
		
		intercept&-0.13(-0.3,0.02)&-0.23(-0.44,0)&-0.21(-0.43,0.01)&-0.15(-0.42,0.05)&-0.17(-0.57,0.06)\\    
		slope&0.03(-0.1,0.2)&-0.12(-0.35,0.08)&-0.13(-0.35,0.07)&-0.14(-0.91,0.38)&-0.24(-1.41,0.41)\\       
		variance&0.01(0,0.04)&0.02(0,0.08)&0.02(0,0.09)&0.01(0,0.04)&0.01(0,0.04)\\                          
		$R^{2}_{adjusted}$&0.32(0,0.95)&0.42(0,0.95)&0.47(0,0.97)&0.32(0,0.95)&0.35(0,0.96)\\   
		
		\multicolumn{6}{c}{\emph{Anti-angiogenic agents }} \\    
		
		intercept&0.03(-0.11,0.2)&0.03(-0.2,0.25)&0.03(-0.19,0.25)&0.02(-0.1,0.14)&0.02(-0.1,0.14)\\         
		slope&-0.33(-0.91,0.09)&-0.86(-1.63,-0.25)&-0.85(-1.65,-0.24)&0.37(-0.01,0.79)&0.35(-0.03,0.78)\\    
		variance&0.02(0,0.07)&0.04(0,0.16)&0.03(0,0.14)&0.02(0,0.06)&0.01(0,0.06)\\                          
		$R^{2}_{adjusted}$&0.5(0,0.97)&0.73(0.09,0.99)&0.72(0.09,0.99)&0.58(0.01,0.97)&0.54(0.01,0.96)\\

		\hline	
		
	\end{tabular}
	\caption{Surrogacy criteria: posterior means, between study variances and associated 95\% Credible Intervals obtained from the bivariate and structured trivariate models, applied to data where all three outcomes are reported for each trial (33 trial)-Prior for between-studies correlation: U(-1,1);  surrogate outcomes TR: Treatment response rate, PFS: Progression free survival and final outcome OS: Overall survival}
	\label{sensrmprior}
\end{table}

\section{Results of alternative trivariate model parameterisations}

A sensitivity analysis to the structure of the between-studies variance-covariance matrix, was carried out. In the main analysis, the between-studies model (in the product normal formulation described by Eq. (2) of the main manuscript), we assumed conditional independence between the true treatment effect on the first surrogate endpoint (TR and the treatment effect on the final outcome (OS). In this sensitivity analysis, two alternative parameterisations were used. In the first one (denoted in Table 10 as 3D structured **) we assumed conditional independence of the true treatment effect on the surrogate endpoints (TR and PFS) whilst the treatment effect on the final outcome was conditional on both treatment effects on the two surrogate endpoints. The second model in sensitivity analysis assumed fully unstructured between-studies variance-covariance matrix, hence the true treatment effects on all three outcomes being correlated.\\

\begin{table}[h!]
	\centering
	\scalebox{0.9}{	\begin{tabular}{ccccccc}	
			&PFS OS &PFS OS  &PFS OS &PFS OS                                      \\
			&2D &3D structured$^{*}$ & 3D structured$^{**}$ & 3D unstructured\\	
			\hline
			intercept&-0.02(-0.06,0.03)&-0.02(-0.06,0.03)&-0.02(-0.06,0.03)&-0.06(-0.14,0.01)\\                   
			slope&0.22(0.03,0.41)&0.19(0.02,0.4)&0.13(0.01,0.33)&0.29(-0.01,0.73)\\                         
			variance&0(0,0.01)&0(0,0.01)&0(0,0.01)&0(0,0.01)\\                                                 
			$R^{2}_{adjusted}$&0.58(0.06,0.97)&0.5(0.02,0.95)&0.35(0,0.87)&0.37(0.04,0.89)\\                   
			DIC&-116.92(-136.5,-95.37)&-117.61(-144.9,-87.1)&-119.96(-146.6,-91.58)&-120.77(-148,90.75)\\
			\hline	
	\end{tabular}}
	\caption{Surrogacy criteria obtained from the bivariate, structured trivariate model$^{*}$ as described in the manuscript (assuming conditional independence between true effects on Response and on OS), the new structured model $^{**}$ (assuming conditional independence between true effects on Response and on PFS) and the unstructured model (no structure between the true effects on surrogate endpoints) applied to the complete dataset (33 studies) }
	\label{dicall}
\end{table}	

\newpage
\section{Reference list: Set of 33 trials reporting all three outcomes; base case scenario 33 studies}

1. Cassidy J, Clarke S, Díaz-Rubio E, Scheithauer W, Figer A, Wong R, et al. Randomized phase III study of capecitabine plus oxaliplatin compared with fluorouracil/folinic acid plus oxaliplatin as first-line therapy for metastatic colorectal cancer.  Journal of clinical oncology; 2008. p. 2006-12.\\

2. Comella P, Massidda B, Filippelli G, Palmeri S, Natale D, Farris A, et al. Ox
aliplatin plus high-dose folinic acid and 5-fluorouracil i.v. bolus (OXAFAFU) versus irinotecan plus high-dose folinic acid and 5-fluorouracil i.v. bolus (IRIFAFU) in patients with metastatic colorectal carcinoma: a Southern Italy Cooperative Oncology Group phase III trial.  2005; 6:[878-86].\\ 

3. Cunningham D, Humblet Y, Siena S, Khayat D, Bleiberg H, Santoro A, et al. Cetuximab monotherapy and cetuximab plus irinotecan in irinotecan-refractory metastatic colorectal cancer.  2004; 4:[337-45]. \\

4. Cunningham D, Sirohi B, Pluzanska A, Utracka-Hutka B, Zaluski J, Glynne-Jones R, et al. Two different first-line 5-fluorouracil regimens with or without oxaliplatin in patients with metastatic colorectal cancer.  2009; 2:[244-50]. \\

5. Diaz-Rubio E, Gomez-Espana A, Massuti B, Sastre J, Abad A, Valladares M, et al. First-Line XELOX plus bevacizumab followed by XELOX plus bevacizumab or single-agent bevacizumab as maintenance therapy in patients with metastatic colorectal cancer: The phase III MACRO TTD study.  2012; 2:[96-103].\\

6. Douillard JY, Siena S, Cassidy J, Tabernero J, Burkes R, Barugel M, et al. Randomized, phase III trial of panitumumab with infusional fluorouracil, leucovorin, and oxaliplatin (FOLFOX4) versus FOLFOX4 alone as first-line treatment in patients with previously untreated metastatic colorectal cancer: the PRIME study.  2010; 31:[4697-705]. \\

7. Ducreux M, Bennouna J, Hebbar M, Ychou M, Lledo G, Conroy T, et al. Capecitabine plus oxaliplatin (XELOX) versus 5-fluorouracil/leucovorin plus oxaliplatin (FOLFOX-6) as first-line treatment for metastatic colorectal cancer.  2011; 3:[682-90]. \\

8. Ducreux M, Malka D, Mendiboure J, Etienne PL, Texereau P, Auby D, et al. Sequential versus combination chemotherapy for the treatment of advanced colorectal cancer (FFCD 2000-05): An open-label, randomised, phase 3 trial.  2011; 11:[1032-44]. \\

9. Fischer von Weikersthal L, Schalhorn A, Stauch M, Quietzsch D, Maubach PA, Lambertz H, et al. Phase III trial of irinotecan plus infusional 5-fluorouracil/folinic acid versus irinotecan plus oxaliplatin as first-line treatment of advanced colorectal cancer.  2011; 2:[206-14]. \\

10. Glimelius B, Sorbye H, Balteskard L, Bystrom P, Pfeiffer P, Tveit K, et al. A randomized phase III multicenter trial comparing irinotecan in combination with the Nordic bolus 5-FU and folinic acid schedule or the bolus/infused de Gramont schedule (Lv5FU2) in patients with metastatic colorectal cancer.  2008; 5:[909-14]. \\

11. Guan ZZ, Xu JM, Luo RC, Feng FY, Wang LW, Shen L, et al. Efficacy and safety of bevacizumab plus chemotherapy in chinese patients with metastatic colorectal cancer:A randomized phase iii artist trial.  2011; 10:[682-9]. \\

12. Hecht JR, Mitchell E, Chidiac T, Scroggin C, Hagenstad C, Spigel D, et al. A randomized phase IIIB trial of chemotherapy, bevacizumab, and panitumumab compared with chemotherapy and bevacizumab alone for metastatic colorectal cancer.  2009; 5:[672-80]. \\

13. Hoff PM, Hochhaus A, Pestalozzi BC, Tebbutt NC, Li J, Kim TW, et al. Cediranib plus FOLFOX/CAPOX versus placebo plus FOLFOX/CAPOX in patients with previously untreated metastatic colorectal cancer: a randomized, double-blind, phase III study (HORIZON II).  2012;29:[3596-603]. \\

14. Jonker DJ, O'Callaghan CJ, Karapetis CS, Zalcberg JR, Tu D, Au HJ, et al. Cetuximab for the treatment of colorectal cancer.  2007; 20:[2040-8].\\

15. Kabbinavar FF, Hambleton J, Mass RD, Hurwitz HI, Bergsland E, Sarkar S. Combined analysis of efficacy: the addition of bevacizumab to fluorouracil/leucovorin improves survival for patients with metastatic colorectal cancer.  Journal of clinical oncology; 2005. p. 3706-12.\\

16. Kerr DJ, McArdle CS, Ledermann J, Taylor I, Sherlock DJ, Schlag PM, et al. Intrahepatic arterial versus intravenous fluorouracil and folinic acid for colorectal cancer liver metastases: a multicentre randomised trial.  2003; 9355:[368-73]. \\

17. Köhne CH, Wils J, Lorenz M, Schöffski P, Voigtmann R, Bokemeyer C, et al. Randomized phase III study of high-dose fluorouracil given as a weekly 24-hour infusion with or without leucovorin versus bolus fluorouracil plus leucovorin in advanced colorectal cancer: European organization of Research and Treatment of Cancer Gastrointestinal Group Study 40952.  Journal of clinical oncology; 2003. p. 3721-8.\\

18. Koopman M, Antonini NF, Douma J, Wals J, Honkoop AH, Erdkamp FL, et al. Sequential versus combination chemotherapy with capecitabine, irinotecan, and oxaliplatin in advanced colorectal cancer (CAIRO): a phase III randomised controlled trial.  Lancet; 2007. p. 135-42.\\

19. Labianca R, Sobrero A, Isa L, Cortesi E, Barni S, Nicolella D, et al. Intermittent versus continuous chemotherapy in advanced colorectal cancer: A randomised 'GISCAD' trial.  2011; 5:[1236-42]. \\

20. Masi G, Vasile E, Loupakis F, Cupini S, Fornaro L, Baldi G, et al. Randomized trial of two induction chemotherapy regimens in metastatic colorectal cancer: an updated analysis.  2011; 1:[21-30]. \\

21. Peeters M, Price TJ, Cervantes A, Sobrero AF, Ducreux M, Hotko Y, et al. Randomized phase III study of panitumumab with fluorouracil, leucovorin, and irinotecan (FOLFIRI) compared with FOLFIRI alone as second-line treatment in patients with metastatic colorectal cancer.  2010; 31:[4706-13]. \\

22. Porschen R, Arkenau HT, Kubicka S, Greil R, Seufferlein T, Freier W, et al. Phase III study of capecitabine plus oxaliplatin compared with fluorouracil and leucovorin plus oxaliplatin in metastatic colorectal cancer: A final report of the AIO colorectal study group.  2007; 27:[4217-23]. \\

23. Rothenberg ML, LaFleur B, Levy DE, Washington MK, Morgan-Meadows SL, Ramanathan RK, et al. Randomized phase II trial of the clinical and biological effects of two dose levels of gefitinib in patients with recurrent colorectal adenocarcinoma.  2005; 36:[9265-74]. \\

24. Schmoll H-J, Cunningham D, Sobrero A, Karapetis CS, Rougier P, Koski SL, et al. Cediranib with mFOLFOX6 versus bevacizumab with mFOLFOX6 as first-line treatment for patients with advanced colorectal cancer: a double-blind, randomized phase III study (HORIZON III).  2012; 29:[3588-95]. \\

25. Sobrero AF, Maurel J, Fehrenbacher L, Scheithauer W, Abubakr YA, Lutz MP, et al. EPIC: phase III trial of cetuximab plus irinotecan after fluoropyrimidine and oxaliplatin failure in patients with metastatic colorectal cancer.  2008; 14:[2311-9]. \\

26. Souglakos J, Ziras N, Kakolyris S, Boukovinas I, Kentepozidis N, Makrantonakis P, et al. Randomised phase-II trial of CAPIRI (capecitabine, irinotecan) plus bevacizumab vs FOLFIRI (folinic acid, 5-fluorouracil, irinotecan) plus bevacizumab as first-line treatment of patients with unresectable/metastatic colorectal cancer (mCRC).  2012; 3:[453-9]. \\

27. Tebbutt NC, Wilson K, Gebski VJ, Cummins MM, Zannino D, van Hazel GA, et al. Capecitabine, bevacizumab, and mitomycin in first-line treatment of metastatic colorectal cancer: results of the Australasian Gastrointestinal Trials Group Randomized Phase III MAX Study.  2010; 19:[3191-8]. \\

28. Tournigand C, Cervantes A, Figer A, Lledo G, Flesch M, Buyse M, et al. OPTIMOX1: A randomized study of FOLFOX4 or FOLFOX7 with oxaliplatin in a stop-and-go fashion in advanced colorectal cancer - A GERCOR study.  2006; 3:[394-400]. \\

29. Tveit KM, Guren T, Glimelius B, Pfeiffer P, Sorbye H, Pyrhonen S, et al. Phase III trial of cetuximab with continuous or intermittent fluorouracil, leucovorin, and oxaliplatin (Nordic FLOX) versus FLOX alone in first-line treatment of metastatic colorectal cancer: the NORDIC-VII study.  2012; 15:[1755-62]. \\

30. Underhill C, Goldstein D, Gorbounova VA, Biakhov MY, Bazin IS, Granov DA, et al. A randomized phase II trial of pemetrexed plus irinotecan (ALIRI) versus leucovorin-modulated 5-FU plus irinotecan (FOLFIRI) in first-line treatment of locally advanced or metastatic colorectal cancer.  2007; 1-2:[9-20]. \\

31. Van Cutsem E, Peeters M, Siena S, Humblet Y, Hendlisz A, Neyns B, et al. Open-label phase III trial of panitumumab plus best supportive-care compared with best supportive-care alone in patients with chemotherapy- refractory metastatic colorectal cancer.  2007; 13:[1658-64]. \\

32. Van Cutsem E, Kohne CH, Lang I, Folprecht G, Nowacki MP, Cascinu S, et al. Cetuximab plus irinotecan, fluorouracil, and leucovorin as first-line treatment for metastatic colorectal cancer: Updated analysis of overall survival according to tumor KRAS and BRAF mutation status.  2011; 15:[2011-9].\\ 

33. Van Cutsem E, Tejpar S, Vanbeckevoort D, Peeters M, Humblet Y, Gelderblom H, et al. Intrapatient cetuximab dose escalation in metastatic colorectal cancer according to the grade of early skin reactions: the randomized EVEREST study.  2012; 23:[2861-8]. 

\newpage

\section{Reference list: Set of 51 trials reporting at least two outcomes}		

1.	Adams RA, Meade AM, Seymour MT, Wilson RH, Madi A, Fisher D, et al. Intermittent versus continuous oxaliplatin and fluoropyrimidine combination chemotherapy for first-line treatment of advanced colorectal cancer: Results of the randomised phase 3 MRC COIN trial.  2011  ; 7:[642-53]. \\

2. Bendell JC, Nemunaitis J, Vukelja SJ, Hagenstad C, Campos LT, Hermann RC, et al. Randomized placebo-controlled phase II trial of perifosine plus capecitabine as second- or third-line therapy in patients with metastatic colorectal cancer.  2011; 33:[4394-400]. \\

3. Cassidy J, Clarke S, Díaz-Rubio E, Scheithauer W, Figer A, Wong R, et al. Randomized phase III study of capecitabine plus oxaliplatin compared with fluorouracil/folinic acid plus oxaliplatin as first-line therapy for metastatic colorectal cancer.  Journal of clinical oncology; 2008. p. 2006-12.\\

4. Chibaudel B, Maindrault-Goebel F, Lledo G, Mineur L, André T, Bennamoun M, et al. Can chemotherapy be discontinued in unresectable metastatic colorectal cancer? The GERCOR OPTIMOX2 Study.  Journal of clinical oncology; 2009. p. 5727-33.\\
5.  Colucci G, Gebbia V, Paoletti G, Giuliani F, Caruso M, Gebbia N, et al. Phase III randomized trial of FOLFIRI versus FOLFOX4 in the treatment of advanced colorectal cancer: A Multicenter Study of the Gruppo Oncologico Dell'Italia Meridionale.  2005; 22:[4866-75]. \\

6. Comella P, Massidda B, Filippelli G, Palmeri S, Natale D, Farris A, et al. Ox
aliplatin plus high-dose folinic acid and 5-fluorouracil i.v. bolus (OXAFAFU) versus irinotecan plus high-dose folinic acid and 5-fluorouracil i.v. bolus (IRIFAFU) in patients with metastatic colorectal carcinoma: a Southern Italy Cooperative Oncology Group phase III trial.  2005; 6:[878-86].\\
 
7. Cunningham D, Humblet Y, Siena S, Khayat D, Bleiberg H, Santoro A, et al. Cetuximab monotherapy and cetuximab plus irinotecan in irinotecan-refractory metastatic colorectal cancer.  2004; 4:[337-45]. \\

8. Cunningham D, Sirohi B, Pluzanska A, Utracka-Hutka B, Zaluski J, Glynne-Jones R, et al. Two different first-line 5-fluorouracil regimens with or without oxaliplatin in patients with metastatic colorectal cancer.  2009; 2:[244-50]. \\

9. Diaz-Rubio E, Tabernero J, Gomez-Espana A, Massuti B, Sastre J, Chaves M, et al. Phase III study of capecitabine plus oxaliplatin compared with continuous-infusion fluorouracil plus oxaliplatin as first-line therapy in metastatic colorectal cancer: Final report of the Spanish Cooperative Group for the treatment of digestive tumors trial.  2007; 27:[4224-30]. \\

10. Diaz-Rubio E, Gomez-Espana A, Massuti B, Sastre J, Abad A, Valladares M, et al. First-Line XELOX plus bevacizumab followed by XELOX plus bevacizumab or single-agent bevacizumab as maintenance therapy in patients with metastatic colorectal cancer: The phase III MACRO TTD study.  2012; 2:[96-103].\\
11. Douillard JY, Siena S, Cassidy J, Tabernero J, Burkes R, Barugel M, et al. Randomized, phase III trial of panitumumab with infusional fluorouracil, leucovorin, and oxaliplatin (FOLFOX4) versus FOLFOX4 alone as first-line treatment in patients with previously untreated metastatic colorectal cancer: the PRIME study.  2010; 31:[4697-705]. \\

12. Ducreux M, Bennouna J, Hebbar M, Ychou M, Lledo G, Conroy T, et al. Capecitabine plus oxaliplatin (XELOX) versus 5-fluorouracil/leucovorin plus oxaliplatin (FOLFOX-6) as first-line treatment for metastatic colorectal cancer.  2011; 3:[682-90]. \\

13. Ducreux M, Malka D, Mendiboure J, Etienne PL, Texereau P, Auby D, et al. Sequential versus combination chemotherapy for the treatment of advanced colorectal cancer (FFCD 2000-05): An open-label, randomised, phase 3 trial.  2011; 11:[1032-44]. \\

14. Fischer von Weikersthal L, Schalhorn A, Stauch M, Quietzsch D, Maubach PA, Lambertz H, et al. Phase III trial of irinotecan plus infusional 5-fluorouracil/folinic acid versus irinotecan plus oxaliplatin as first-line treatment of advanced colorectal cancer.  2011; 2:[206-14]. \\

15. Fuchs CS, Marshall J, Mitchell E, Wierzbicki R, Ganju V, Jeffery M, et al. Randomized, controlled trial of irinotecan plus infusional, bolus, or oral fluoropyrimidines in first-line treatment of metastatic colorectal cancer: results from the BICC-C Study.  Journal of clinical oncology; 2007. p. 4779-86.\\

16. Giacchetti S, Bjarnason G, Garufi C, Genet D, Iacobelli S, Tampellini M, et al. Phase III trial comparing 4-day chronomodulated therapy versus 2-day conventional delivery of fluorouracil, leucovorin, and oxaliplatin as first-line chemotherapy of metastatic colorectal cancer: the European Organisation for Research and Treatment of Cancer Chronotherapy Group.  2006; 22:[3562-9]. \\

17. Glimelius B, Sorbye H, Balteskard L, Bystrom P, Pfeiffer P, Tveit K, et al. A randomized phase III multicenter trial comparing irinotecan in combination with the Nordic bolus 5-FU and folinic acid schedule or the bolus/infused de Gramont schedule (Lv5FU2) in patients with metastatic colorectal cancer.  2008; 5:[909-14]. \\

18. Goldberg RM, Sargent DJ, Morton RF, Fuchs CS, Ramanathan RK, Williamson SK, et al. Randomized controlled trial of reduced-dose bolus fluorouracil plus leucovorin and irinotecan or infused fluorouracil plus leucovorin and oxaliplatin in patients with previously untreated metastatic colorectal cancer: a North American Intergroup Trial.  2006; 21:[3347-53]. \\

19. Guan ZZ, Xu JM, Luo RC, Feng FY, Wang LW, Shen L, et al. Efficacy and safety of bevacizumab plus chemotherapy in chinese patients with metastatic colorectal cancer:A randomized phase iii artist trial.  2011; 10:[682-9]. \\

20. Haller DG, Rothenberg ML, Wong AO, Koralewski PM, Miller Jr WH, Bodoky G, et al. Oxaliplatin plus irinotecan compared with irinotecan alone as second-line treatment after single-agent fluoropyrimidine therapy for metastatic colorectal carcinoma.  2008; 28:[4544-50]. \\

21. Hecht JR, Mitchell E, Chidiac T, Scroggin C, Hagenstad C, Spigel D, et al. A randomized phase IIIB trial of chemotherapy, bevacizumab, and panitumumab compared with chemotherapy and bevacizumab alone for metastatic colorectal cancer.  2009; 5:[672-80]. \\

22. Hecht JR, Trarbach T, Hainsworth JD, Major P, Jager E, Wolff RA, et al. Randomized, placebo-controlled, phase III study of first-line oxaliplatin-based chemotherapy plus PTK787/ZK 222584, an oral vascular endothelial growth factor receptor inhibitor, in patients with metastatic colorectal adenocarcinoma.  2011; 15:[1997-2003]. \\

23. Hoff PM, Hochhaus A, Pestalozzi BC, Tebbutt NC, Li J, Kim TW, et al. Cediranib plus FOLFOX/CAPOX versus placebo plus FOLFOX/CAPOX in patients with previously untreated metastatic colorectal cancer: a randomized, double-blind, phase III study (HORIZON II).  2012;29:[3596-603]. \\

24. Jonker DJ, O'Callaghan CJ, Karapetis CS, Zalcberg JR, Tu D, Au HJ, et al. Cetuximab for the treatment of colorectal cancer.  2007; 20:[2040-8].

25. Kabbinavar FF, Hambleton J, Mass RD, Hurwitz HI, Bergsland E, Sarkar S. Combined analysis of efficacy: the addition of bevacizumab to fluorouracil/leucovorin improves survival for patients with metastatic colorectal cancer.  Journal of clinical oncology; 2005. p. 3706-12.\\

26. Kalofonos HP, Papakostas P, Makatsoris T, Papamichael D, Vourli G, Xanthakis I, et al. Irinotecan/fluorouracil/\\ leucovorin or the same regimen followed by oxaliplatin/fluorouracil/leucovorin in metastatic colorectal cancer.  2010; 10:[4325-33]. \\

27. Kerr DJ, McArdle CS, Ledermann J, Taylor I, Sherlock DJ, Schlag PM, et al. Intrahepatic arterial versus intravenous fluorouracil and folinic acid for colorectal cancer liver metastases: a multicentre randomised trial.  2003; 9355:[368-73]. \\

28. Kim GP, Sargent DJ, Mahoney MR, Rowland KM, Jr., Philip PA, Mitchell E, et al. Phase III noninferiority trial comparing irinotecan with oxaliplatin, fluorouracil, and leucovorin in patients with advanced colorectal carcinoma previously treated with fluorouracil: N9841.  2009; 17:[2848-54]. \\

29. Köhne CH, Wils J, Lorenz M, Schöffski P, Voigtmann R, Bokemeyer C, et al. Randomized phase III study of high-dose fluorouracil given as a weekly 24-hour infusion with or without leucovorin versus bolus fluorouracil plus leucovorin in advanced colorectal cancer: European organization of Research and Treatment of Cancer Gastrointestinal Group Study 40952.  Journal of clinical oncology; 2003. p. 3721-8.\\

30. Koopman M, Antonini NF, Douma J, Wals J, Honkoop AH, Erdkamp FL, et al. Sequential versus combination chemotherapy with capecitabine, irinotecan, and oxaliplatin in advanced colorectal cancer (CAIRO): a phase III randomised controlled trial.  Lancet; 2007. p. 135-42.\\

31. Labianca R, Sobrero A, Isa L, Cortesi E, Barni S, Nicolella D, et al. Intermittent versus continuous chemotherapy in advanced colorectal cancer: A randomised 'GISCAD' trial.  2011; 5:[1236-42]. \\

32. Masi G, Vasile E, Loupakis F, Cupini S, Fornaro L, Baldi G, et al. Randomized trial of two induction chemotherapy regimens in metastatic colorectal cancer: an updated analysis.  2011; 1:[21-30]. \\

33. Maughan TS, Adams RA, Smith CG, Meade AM, Seymour MT, Wilson RH, et al. Addition of cetuximab to oxaliplatin-based first-line combination chemotherapy for treatment of advanced colorectal cancer: results of the randomised phase 3 MRC COIN trial.  Lancet; 2011. p. 2103-14.\\

34. Ocvirk J, Brodowicz T, Wrba F, Ciuleanu TE, Kurteva G, Beslija S, et al. Cetuximab plus FOLFOX6 or FOLFIRI in metastatic colorectal cancer: CECOG trial.  2010; 25:[3133-43]. \\

35. Peeters M, Price TJ, Cervantes A, Sobrero AF, Ducreux M, Hotko Y, et al. Randomized phase III study of panitumumab with fluorouracil, leucovorin, and irinotecan (FOLFIRI) compared with FOLFIRI alone as second-line treatment in patients with metastatic colorectal cancer.  2010; 31:[4706-13]. \\

36. Porschen R, Arkenau HT, Kubicka S, Greil R, Seufferlein T, Freier W, et al. Phase III study of capecitabine plus oxaliplatin compared with fluorouracil and leucovorin plus oxaliplatin in metastatic colorectal cancer: A final report of the AIO colorectal study group.  2007; 27:[4217-23]. \\

37. Rothenberg ML, LaFleur B, Levy DE, Washington MK, Morgan-Meadows SL, Ramanathan RK, et al. Randomized phase II trial of the clinical and biological effects of two dose levels of gefitinib in patients with recurrent colorectal adenocarcinoma.  2005; 36:[9265-74]. \\

38. Schmoll H-J, Cunningham D, Sobrero A, Karapetis CS, Rougier P, Koski SL, et al. Cediranib with mFOLFOX6 versus bevacizumab with mFOLFOX6 as first-line treatment for patients with advanced colorectal cancer: a double-blind, randomized phase III study (HORIZON III).  2012; 29:[3588-95]. \\

39. Sobrero AF, Maurel J, Fehrenbacher L, Scheithauer W, Abubakr YA, Lutz MP, et al. EPIC: phase III trial of cetuximab plus irinotecan after fluoropyrimidine and oxaliplatin failure in patients with metastatic colorectal cancer.  2008; 14:[2311-9]. \\

40. Souglakos J, Androulakis N, Syrigos K, Polyzos A, Ziras N, Athanasiadis A, et al. FOLFOXIRI (folinic acid, 5-fluorouracil, oxaliplatin and irinotecan) vs FOLFIRI (folinic acid, 5-fluorouracil and irinotecan) as first-line treatment in metastatic colorectal cancer (MCC): a multicentre randomised phase III trial from the Hellenic Oncology Research Group (HORG).  2006; 6:[798-805]. \\

41. Souglakos J, Ziras N, Kakolyris S, Boukovinas I, Kentepozidis N, Makrantonakis P, et al. Randomised phase-II trial of CAPIRI (capecitabine, irinotecan) plus bevacizumab vs FOLFIRI (folinic acid, 5-fluorouracil, irinotecan) plus bevacizumab as first-line treatment of patients with unresectable/metastatic colorectal cancer (mCRC).  2012; 3:[453-9]. \\

42. Tebbutt NC, Wilson K, Gebski VJ, Cummins MM, Zannino D, van Hazel GA, et al. Capecitabine, bevacizumab, and mitomycin in first-line treatment of metastatic colorectal cancer: results of the Australasian Gastrointestinal Trials Group Randomized Phase III MAX Study.  2010; 19:[3191-8]. \\

43. Tol J, Koopman M, Cats A, Rodenburg CJ, Creemers GJM, Schrama JG, et al. Chemotherapy, bevacizumab, and cetuximab in metastatic colorectal cancer.  2009; 6:[563-72]. \\

44. Tournigand C, Cervantes A, Figer A, Lledo G, Flesch M, Buyse M, et al. OPTIMOX1: A randomized study of FOLFOX4 or FOLFOX7 with oxaliplatin in a stop-and-go fashion in advanced colorectal cancer - A GERCOR study.  2006; 3:[394-400]. \\

45. Tveit KM, Guren T, Glimelius B, Pfeiffer P, Sorbye H, Pyrhonen S, et al. Phase III trial of cetuximab with continuous or intermittent fluorouracil, leucovorin, and oxaliplatin (Nordic FLOX) versus FLOX alone in first-line treatment of metastatic colorectal cancer: the NORDIC-VII study.  2012; 15:[1755-62]. \\

46. Underhill C, Goldstein D, Gorbounova VA, Biakhov MY, Bazin IS, Granov DA, et al. A randomized phase II trial of pemetrexed plus irinotecan (ALIRI) versus leucovorin-modulated 5-FU plus irinotecan (FOLFIRI) in first-line treatment of locally advanced or metastatic colorectal cancer.  2007; 1-2:[9-20]. \\

47. Van Cutsem E, Peeters M, Siena S, Humblet Y, Hendlisz A, Neyns B, et al. Open-label phase III trial of panitumumab plus best supportive-care compared with best supportive-care alone in patients with chemotherapy- refractory metastatic colorectal cancer.  2007; 13:[1658-64]. \\

48. Van Cutsem E, Bajetta E, Valle J, Kohne CH, Hecht JR, Moore M, et al. Randomized, placebo-controlled, phase III study of oxaliplatin, fluorouracil, and leucovorin with or without PTK787/ZK 222584 in patients with previously treated metastatic colorectal adenocarcinoma.  2011; 15:[2004-10]. \\

49. Van Cutsem E, Kohne CH, Lang I, Folprecht G, Nowacki MP, Cascinu S, et al. Cetuximab plus irinotecan, fluorouracil, and leucovorin as first-line treatment for metastatic colorectal cancer: Updated analysis of overall survival according to tumor KRAS and BRAF mutation status.  2011; 15:[2011-9].\\ 

51. Van Cutsem E, Tejpar S, Vanbeckevoort D, Peeters M, Humblet Y, Gelderblom H, et al. Intrapatient cetuximab dose escalation in metastatic colorectal cancer according to the grade of early skin reactions: the randomized EVEREST study.  2012; 23:[2861-8].
100. Wolff RA, Fuchs M, Di Bartolomeo M, Hossain AM, Stoffregen C, Nicol S, et al. A double-blind, randomized, placebo-controlled, phase 2 study of maintenance enzastaurin with 5-fluorouracil/leucovorin plus bevacizumab after first-line therapy for metastatic colorectal cancer.  2012; 17:[4132-8].

\end{document}